\documentclass[11pt]{article}
\pdfoutput=1

\usepackage{chngpage}

\usepackage{amsmath,amsfonts,enumerate,array
}

\usepackage{graphicx}
\usepackage[usenames]{color}
\usepackage[english]{babel}

\usepackage{fancybox}

\usepackage{mciteplus}

\usepackage[
      colorlinks=true,
      linkcolor=black,
      urlcolor=blue,
      citecolor=blue,
      filecolor=blue,
      pdfstartview=FitH,
      pdftitle={},
      pdfauthor={},
      pdfsubject={},
      pdfkeywords={},
      pdfpagemode=UseNone,
      bookmarksopen=true
      ]{hyperref}
\usepackage[all]{hypcap}     

\begin{document}

\numberwithin{equation}{section}


\mathchardef\mhyphen="2D


\newcommand{\be}{\begin{equation}}
\newcommand{\ee}{\end{equation}}
\newcommand{\bea}{\begin{eqnarray}\displaystyle}
\newcommand{\eea}{\end{eqnarray}}
\newcommand{\nnm}{\nonumber}
\newcommand{\nn}{\nonumber}

\def\eq#1{(\ref{#1})}
\newcommand{\secn}[1]{Section~\ref{#1}}

\newcommand{\tbl}[1]{Table~\ref{#1}}
\newcommand{\fig}{Fig.~\ref}

\def\beq{\begin{equation}}
\def\eeq{\end{equation}}
\def\beqa{\begin{eqnarray}}
\def\eeqa{\end{eqnarray}}
\def\bet{\begin{tabular}}
\def\eet{\end{tabular}}
\def\bs{\begin{split}}
\def\es{\end{split}}


\def\a{\alpha}  \def\b{\beta}   \def\c{\chi}    
\def\g{\gamma}  \def\G{\Gamma}  \def\e{\epsilon}  
\def\vep{\varepsilon}   \def\tvep{\widetilde{\varepsilon}}
\def\f{\phi}    \def\F{\Phi}  \def\fb{{\ov \phi}}
\def\vf{\varphi}  \def\m{\mu}  \def\mub{\ov \mu}
\def\n{\nu}  \def\nub{\ov \nu}  \def\o{\omega}
\def\O{\Omega}  \def\r{\rho}  \def\k{\kappa}
\def\kab{\ov \kappa}  \def\s{\sigma}
\def\t{\tau}  \def\th{\theta}  \def\sb{\ov\sigma}  \def\S{\Sigma}
\def\l{\lambda}  \def\L{\Lambda}  \def\p{\psi}


\def\cA{{\cal A}} \def\cB{{\cal B}} \def\cC{{\cal C}}
\def\cD{{\cal D}} \def\cE{{\cal E}} \def\cF{{\cal F}}
\def\cG{{\cal G}} \def\cH{{\cal H}} \def\cI{{\cal I}}
\def\cJ{{\cal J}} \def\cK{{\cal K}} \def\cL{{\cal L}}
\def\cM{{\cal M}} \def\cN{{\cal N}} \def\cO{{\cal O}}
\def\cP{{\cal P}} \def\cQ{{\cal Q}} \def\cR{{\cal R}}
\def\cS{{\cal S}} \def\cT{{\cal T}} \def\cU{{\cal U}}
\def\cV{{\cal V}} \def\cW{{\cal W}} \def\cX{{\cal X}}
\def\cY{{\cal Y}} \def\cZ{{\cal Z}}

\def\mC{\mathbb{C}} \def\mP{\mathbb{P}}  
\def\mR{\mathbb{R}} \def\mZ{\mathbb{Z}} 
\def\mT{\mathbb{T}} \def\mN{\mathbb{N}}
\def\mH{\mathbb{H}} \def\mX{\mathbb{X}}

\def\CP{\mathbb{CP}}
\def\RP{\mathbb{RP}}
\def\Z{\mathbb{Z}}
\def\N{\mathbb{N}}
\def\H{\mathbb{H}}

\newcommand{\rmd}{\mathrm{d}}
\newcommand{\rmx}{\mathrm{x}}

\def\tA{ {\widetilde A} } 

\def\one{{\hbox{\kern+.5mm 1\kern-.8mm l}}}
\def\zero{{\hbox{0\kern-1.5mm 0}}}


\newcommand{\bra}[1]{{\langle {#1} |\,}}
\newcommand{\ket}[1]{{\,| {#1} \rangle}}
\newcommand{\braket}[2]{\ensuremath{\langle #1 | #2 \rangle}}
\newcommand{\Braket}[2]{\ensuremath{\langle\, #1 \,|\, #2 \,\rangle}}
\newcommand{\norm}[1]{\ensuremath{\left\| #1 \right\|}}
\def\corr#1{\left\langle \, #1 \, \right\rangle}
\def\vac{|0\rangle}


\def\d{ \partial } 
\def\zb{{\bar z}}

\newcommand{\sq}{\square}
\newcommand{\IP}[2]{\ensuremath{\langle #1 , #2 \rangle}}    

\newcommand{\floor}[1]{\left\lfloor #1 \right\rfloor}
\newcommand{\ceil}[1]{\left\lceil #1 \right\rceil}

\newcommand{\dbyd}[1]{\ensuremath{ \frac{\d}{\d {#1}}}}
\newcommand{\ddbyd}[1]{\ensuremath{ \frac{\d^2}{\d {#1}^2}}}

\newcommand{\Zd}{\ensuremath{ Z^{\dagger}}}
\newcommand{\Xd}{\ensuremath{ X^{\dagger}}}
\newcommand{\Ad}{\ensuremath{ A^{\dagger}}}
\newcommand{\Bd}{\ensuremath{ B^{\dagger}}}
\newcommand{\Ud}{\ensuremath{ U^{\dagger}}}
\newcommand{\Td}{\ensuremath{ T^{\dagger}}}

\newcommand{\T}[3]{\ensuremath{ #1{}^{#2}_{\phantom{#2} \! #3}}}		

\newcommand{\tr}{\operatorname{tr}}
\newcommand{\sech}{\operatorname{sech}}
\newcommand{\Spin}{\operatorname{Spin}}
\newcommand{\Sym}{\operatorname{Sym}}
\newcommand{\Com}{\operatorname{Com}}
\def\adj{\textrm{adj}}
\def\id{\textrm{id}}

\def\ha{\frac{1}{2}}
\def\tha{\tfrac{1}{2}}
\def\wt{\widetilde}
\def\ra{\rangle}
\def\la{\langle}

\def\pb{\ov\psi}
\def\pt{\widetilde{\psi}}
\def\at{\widetilde{\a}}
\def\cb{\ov\chi}
\def\d{\partial}
\def\db{\bar\partial}
\def\delb{\bar\partial}
\def\dbar{\ov\partial}
\def\dag{\dagger}
\def\dalpha{{\dot\alpha}}
\def\dbeta{{\dot\beta}}
\def\dgamma{{\dot\gamma}}
\def\ddelta{{\dot\delta}}
\def\ad{{\dot\alpha}}
\def\bd{{\dot\beta}}
\def\dg{{\dot\gamma}}
\def\dd{{\dot\delta}}
\def\th{\theta}
\def\Th{\Theta}
\def\eb{{\ov \epsilon}}
\def\gb{{\ov \gamma}}
\def\wb{{\ov w}}
\def\Wb{{\ov W}}
\def\D{\Delta}
\def\DD{\Delta^\dag}
\def\Db{\ov D}

\def\ov{\overline}
\def\Slash{\, / \! \! \! \!}
\def\dslash{\partial\!\!\!/} 
\def\Dslash{D\!\!\!\!/\,\,}
\def\fslash#1{\slash\!\!\!#1}
\def\Fslash#1{\slash\!\!\!\!#1}

\def\del{\partial}
\def\delb{\bar\partial}
\newcommand{\ex}[1]{{\rm e}^{#1}} 
\def\ii{{i}}

\newcommand{\vs}[1]{\vspace{#1 mm}}

\newcommand{\ve}{{\vec{\e}}}
\newcommand{\shalf}{\frac{1}{2}}

\newcommand{\lb}{\rangle}
\newcommand{\al}{\ensuremath{\alpha'}}
\newcommand{\ap}{\ensuremath{\alpha'}}

\newcommand{\bean}{\begin{eqnarray*}}
\newcommand{\eean}{\end{eqnarray*}}
\newcommand{\ft}[2]{{\textstyle {\frac{#1}{#2}} }}

\newcommand{\hsp}{\hspace{0.5cm}}
\def\half{{\textstyle{1\over2}}}
\let\ci=\cite \let\re=\ref
\let\se=\section \let\sse=\subsection \let\ssse=\subsubsection

\newcommand{\dpb}{D$p$-brane}
\newcommand{\dpbs}{D$p$-branes}

\def\gh{{\rm gh}}
\def\sgh{{\rm sgh}}
\def\NS{{\rm NS}}
\def\R{{\rm R}}
\def\Qp{Q_{\rm P}}
\def\QP{Q_{\rm P}}

\newcommand\dott[2]{#1 \! \cdot \! #2}

\def\eo{\overline{e}}



\def\p{\partial}
\def\h{{1\over 2}}

\def\d{\partial}
\def\la{\lambda}
\def\eps{\epsilon}
\def\bb{\bigskip}
\def\tg{\widetilde\gamma}
\newcommand{\dm}{\begin{displaymath}}
\newcommand{\edm}{\end{displaymath}}
\renewcommand{\b}{\widetilde{B}}
\newcommand{\gm}{\Gamma}
\newcommand{\ac}[2]{\ensuremath{\{ #1, #2 \}}}
\renewcommand{\ell}{l}
\newcommand{\z}{\ell}
\def\bb{$\bullet$}
\def\Qbar{{\bar Q}_1}
\def\QPbar{{\bar Q}_p}

\def\q{\quad}

\def\bn{B_\circ}

\let\a=\alpha \let\b=\beta \let\g=\gamma 
\let\e=\epsilon
\let\c=\chi \let\th=\theta  \let\k=\kappa
\let\l=\lambda \let\m=\mu \let\n=\nu \let\x=\xi \let\r=\rho
\let\s=\sigma 
\let\vp=\varphi \let\vep=\varepsilon
\let\w=\omega  \let\G=\Gamma \let\D=\Delta \let\Th=\Theta \let\P=\Pi \let\S=\Sigma

\let\t=\tilde

\def\h{{1\over 2}}

\def\r{\rightarrow}
\def\Ri{\Rightarrow}

\def\nn{\nonumber\\}
\let\bm=\bibitem
\def\Kt{{\widetilde K}}
\def\b{\bigskip}

\let\p=\partial

\newcommand{\mm}{Balasubramanian:2000rt,*Maldacena:2000dr}
\newcommand{\MM}{Balasubramanian:2000rt,*Maldacena:2000dr}
\newcommand{\cvet}{Cvetic:1996xz,*Cvetic:1997uw}
\newcommand{\lmRot}{Lunin:2001fv}
\newcommand{\GMR}{Gutowski:2003rg}
\newcommand{\lmAdS}{Lunin:2001jy}
\newcommand{\lmm}{Lunin:2002iz}
\newcommand{\CarMcCon}{Cariglia:2004wu}
\newcommand{\Grana}{Schwarz:1983qr,Grana:2000jj}
\newcommand{\sv}{Strominger:1996sh}
\newcommand{\kst}{Kanitscheider:2007wq}
\newcommand{\bgsw}{Bena:2011dd}
\newcommand{\fuzz}{Mathur:2005zp,*Bena:2007kg,*Skenderis:2008qn,*Chowdhury:2010ct,Mathur:2012zp}
\newcommand{\adscft}{Maldacena:1997re,*Gubser:1998bc,*Witten:1998qj}


\begin{flushright}
\end{flushright}
\vspace{15mm}

\begin{center}
{\LARGE Adding momentum to supersymmetric geometries}
\\
\vspace{18mm}
{\bf    Oleg Lunin}${}^{1}$, ~{\bf Samir D. Mathur}${}^{2}$, ~{\bf David Turton}${}^{2}$

\vspace{6mm}

${}^{1}$Department of Physics,\\ University at Albany (SUNY),\\ Albany, NY 12222, USA\\ 

\vspace{4mm}

${}^{2}$Department of Physics,\\ The Ohio State University,\\ Columbus,
OH 43210, USA\\ 
\vskip 6 mm
olunin@albany.edu\\
mathur.16@osu.edu\\
turton.7@osu.edu
\vspace{12mm}

\end{center}

\begin{abstract}

\vspace{4mm}

\noindent
We consider general supersymmetric solutions to minimal supergravity in six dimensions, trivially lifted to IIB supergravity. To any such solution we add a travelling-wave deformation involving the additional directions. The deformed solution is given in terms of a function which is harmonic in the background geometry. We also present a family of explicit examples describing microstates of the D1-D5 system on $T^4$. In the case where the background contains a large AdS region, the deformation is identified as corresponding to an action of a U(1) current of the D1-D5 orbifold CFT on a given state.

\end{abstract}

\thispagestyle{empty}

\newpage

\setcounter{tocdepth}{1}
\tableofcontents

\baselineskip=15pt
\parskip=3pt

\section{Introduction}
\label{intr}

Supersymmetric solutions to supergravity play an important role in string theory, notably in the AdS/CFT correspondence~\cite{\adscft} and in applications to black hole physics~\cite{Strominger:1996sh}. Solutions which are smooth and horizonless are particularly interesting because they can describe individual microstates of supersymmetric black holes. The gravity description of black hole microstates offers a resolution to the black hole information paradox~\cite{Hawking:1976ra}, known as the fuzzball proposal~\cite{\fuzz}.

Assuming the existence of a Killing spinor, the general form of supersymmetric solutions to certain supergravity theories has been constructed. 
For example, this has been carried out for minimal $\cN=2$ supergravity in 4D~\cite{Tod:1983pm}, minimal supergravity in 5D~\cite{Gauntlett:2002nw} and minimal supergravity in 6D \cite{Gutowski:2003rg}. In the 6D case, considered by Gutowski, Martelli and Reall (GMR), the solution is given in terms of a 2D fiber over a 4D almost hyperkahler base.

Microstates of the 2-charge extremal D1-D5 black hole can be put in GMR form, with the base being flat 4D space. A microstate of the 3-charge D1-D5-P extremal hole was found in \cite{Lunin:2004uu} using the GMR ansatz. 
Similar microstates were found in \cite{Giusto:2004id,*Giusto:2004ip} by taking supersymmetric limits of general black hole solutions. The 4D base of these solutions is not positive definite everywhere~\cite{Giusto:2004kj}; the signature flips from $(4,0)$ to $(0,4)$ within some region. The fiber degenerates at the boundary of this region, such that the overall geometry remains regular with signature $(1,5)$ everywhere. This structure was generalized in \cite{Bena:2004de} to an arbitrary number of negative signature domains, adapted to type IIB supergravity,
and the equations cast in a way that they could be solved sequentially as a set of linear equations with sources.
These developments have led to the construction of large classes of supergravity solutions to string theory~\cite{Bena:2005va,*Berglund:2005vb,*Saxena:2005uk,*Balasubramanian:2006gi,*deBoer:2008fk,*Bena:2010gg,*Bena:2011uw,*Giusto:2011fy,*Giusto:2012gt}. 

To address the information paradox, it is useful to relate supergravity solutions to CFT states under AdS/CFT duality. 
This can be done for solutions which have a large AdS region.
The AdS/CFT map is well understood for supersymmetric 2-charge D1-D5 states \cite{Lunin:2001jy,Lunin:2002iz,Taylor:2005db,Kanitscheider:2007wq}. A 3-charge state carrying one unit of momentum was constructed in \cite{Mathur:2003hj}; this can be mapped to a CFT state generated by a twist operator on the CFT vacuum. The states of \cite{Lunin:2004uu,Giusto:2004id,*Giusto:2004ip} are related to 2-charge states by spectral flow, and so their CFT duals are also known. Finding the CFT interpretation of more general three-charge geometries such as those constructed in \cite{Bena:2004de} remains an open problem.

In this paper we present a deformation which adds a travelling wave to a given background. Our construction applies to a general solution in GMR form, trivially lifted to 10D, 
and the deformation involves the additional directions. 

In more detail, we consider the D1-D5 system on $T^4$, for which the CFT is a deformation of an orbifold CFT~\cite{Strominger:1996sh,Seiberg:1999xz}. At the orbifold point in moduli space, this $\cN=4$ CFT is described by four free bosons $X^\a$ and their superpartners. The CFT has a chiral symmetry algebra whose bosonic generators are the  Virasoro generators $L_n$, the SU(2) R-symmetry currents ${\cal J}^a_n$ and four U(1) currents $J^{\a}$ arising from 
the translation symmetries on $T^4$. The U(1) currents are given by $J^{\a}=\p X^\a$. 

Given a state $|\psi\rangle$ in the D1-D5 orbifold CFT, one can apply the chiral algebra generators
$L_n, {\cal J}^a_n, J^{\a}_n\,$. Doing so adds energy and left-moving momentum, producing a 3-charge D1-D5-P state. One can then ask for the gravity description of the new state. One can also ask the same question for the corresponding coherent states, e.g.~$e^{\mu J^\a_{-n}}|\psi\rangle$.

A simple example is provided by the 2-charge D1-D5 background found in~\cite{\mm}. Following the work of Brown and Henneaux~\cite{Brown:1986nw}, perturbative gravity solutions corresponding to the action of $L_n, {\cal J}^a_n, J^{\a}_n$ on this background were recently constructed in the limit where the added energy was small~\cite{Mathur:2011gz}. The perturbations were found in the approximation scheme used in \cite{Mathur:2003hj}. 

Subsequently, the closed-form linear perturbation describing the U(1) current $J^{\a}_{-n}$ applied to the same 2-charge background was found in~\cite{Mathur:2012tj}. Denoting the background state by $|0\rangle_R$, the perturbation describes the state $J^\a_{-n}|0\rangle_R$. Equivalently, the background and perturbation together describe the coherent state 
$e^{\mu J^\a_{-n}}|0\rangle_R$ in the limit of infinitesimal $\mu$. 

The background solution used in \cite{Mathur:2011gz,Mathur:2012tj} has, in a certain regime of parameters, a large AdS region and flat asymptotics, separated by an intermediate `neck' region.
The deformation arises at the boundary of the AdS region and appears to be related to the `singleton' (or `doubleton') representations that lie at the boundary of AdS \cite{Dirac:1963ta,*Flato:1978qz,*Gunaydin:1986fe}.

In this paper we consider the same U(1) currents $J^{\a}$, but in a much more general context.
We start with various states $|\psi\rangle$ and consider the nonlinear problem, i.e.~we find the gravity description of the coherent states
\bea
e^{\mu J^\a_{-n}}|\psi\rangle \, \nonumber
\eea
with arbitrary $\mu$. For the background states $|\psi\rangle$ we consider the following:

(i) As a first example, we consider the background state $|\psi\rangle = |0\rangle_R$.
In this case we find the nonlinear generalization of the perturbative solution of \cite{Mathur:2012tj}.

(ii) Next we take $|\psi\rangle$ to be described by any solution in GMR form, trivially uplifted to 10D. The set of such solutions includes a class of 2-charge D1-D5 as well as a class of 3-charge D1-D5-P solutions. We find that the state $e^{\mu J^\a_{-n}}|\psi\rangle$ is described by a metric involving a function $\Phi$, which is harmonic in the background geometry describing $|\psi\rangle$. 
Although we emphasize the CFT interpretation, the supergravity deformation may be applied to any trivial uplift of a solution which takes the GMR form.

(iii) As a second example, we present an explicit class of solutions covered in the general treatment (ii). We take $|\psi\rangle$ to describe a D1-D5-P solution of \cite{Lunin:2004uu,Giusto:2004id,*Giusto:2004ip}.
For this case we find the function $\Phi$ explicitly.

Our results demonstrate that nonlinear travelling waves can be added to large classes of supersymmetric supergravity solutions. 
A word of caution is in order, however, regarding solutions with event horizons. 
Travelling waves similar to those constructed in this paper, but on black hole backgrounds, were studied in~\cite{Larsen:1995ss,Cvetic:1995bj,*Tseytlin:1996as,*Tseytlin:1996qg,Horowitz:1996th,*Horowitz:1996cj}. 
It was later found that adding the wave led to curvature singularities at the horizon 
in the form of infinite tidal forces
\cite{Kaloper:1996hr,*Horowitz:1997si} (see also~\cite{Banerjee:2009uk,Jatkar:2009yd}). 
This can be regarded as an instance of the `no-hair theorem': there are no regular perturbations of a black hole horizon. 
In this paper we thus restrict our attention to solutions which are regular and have no event horizons.

\section{The D1-D5 orbifold CFT and the states of interest}
\label{SectCFT}

In this section we review the D1-D5 orbifold CFT
on $T^4$ and the U(1) currents in this theory. We work in type IIB string theory with the compactification
\be
M_{9, 1}~\r~ M_{4, 1}\times S^1\times T^4 \,.
\ee
We consider the bound states of $n_1$ D1 branes wrapping $S^1$ and $n_5$ D5 branes wrapping $S^1\times T^4$. 
We use $t$ for the time coordinate of $M_{4, 1}$ and $y$ for the $S^1$.

In the IR limit, the D1-D5 system is described  by a 1+1 dimensional sigma model. 
The base space of this sigma model is parameterized by $(t,y)$ and the target space is a deformation of the orbifold $(T^4)^N/S_N$, 
where $N=n_1n_5$.
The CFT has $(4,4)$ supersymmetry, and the moduli space preserves this supersymmetry. It is conjectured that in the
moduli space there is an `orbifold point' where the target space is just $(T^4)^N/S_N$ \cite{Seiberg:1999xz}.

The (4,4) CFT with target space $T^4$ has central charge $c=6$ and is described by four real bosons $X^\a$ and their superpartners.
The complete theory with target space $(T^4)^N/S_N$ has $N$ copies of the $c=6$ CFT, with states that are symmetrized between the $N$ copies.  

\subsection{U(1) currents}

We temporarily use $r$ to label different copies of the $c=6$ CFT described above. At the orbifold point, the fields are free. As a result, each copy has four holomorphic U(1) currents
\be
J^{(r),\beta} ~=~ \p X^{(r),\beta}
\ee
arising from translations in the four torus directions. The total CFT has the U(1) currents
\be
J^\beta ~=~ \sum\limits_{r=1}^{N} \p X^{(r),\beta} \,.
\ee
The modes of the current $J^{(r),\beta}$ are then the bosonic oscillator modes $\alpha_{-n}^{(r),\beta}$, and the modes of $J^\beta$ are
\be
J^\beta_{-n} ~=~ \sum\limits_{r=1}^{N} \alpha_{-n}^{(r),\beta} \,.
\ee

\subsection{Background states}

The purpose of this paper is to start with a class of CFT states $|\psi\rangle$ with known gravitational descriptions, and to construct the gravitational description of the state 
\bea \label{eq:J_psi}
e^{\mu J^\a_{-n}}|\psi\rangle \,.
\eea
More generally, one can consider a combination of $J^\a_{-n}$ in different directions $\a$ on the $T^4$,
\bea\label{JonPsi}
\exp{\left(\sum_{n>0} \mu^\alpha_n J^{\a}_{-n} \right)}|\psi\rangle \,.
\eea
We consider the class of CFT states $|\psi\rangle$ whose gravitational descriptions may be obtained by trivial uplifts of solutions to minimal supergravity in six dimensions. We shall discuss this general class in Section \ref{sec:GMRetc}. For now we discuss the particular states which are used in our explicit examples.

First let us consider the particular Ramond sector ground state $|0\rangle_R$ that is obtained by spectral flow
from the NS vacuum $|0\rangle_{NS}$,
\bea \label{eq:0r}
|0\rangle_R &=& \cS |0\rangle_{NS} \,.
\eea
The dual geometry is known, and will be given in \eq{D1D5MM} below. 
In this case the gravitational description of the state \eq{eq:J_psi} is the non-linear extension of the perturbation constructed in \cite{Mathur:2011gz,Mathur:2012tj}, and will be given in (\ref{MMConj}).

A more general class of Ramond ground states can be constructed by performing spectral flow by more than one unit\footnote{Spectral flow by an even number of units brings the vacuum $|0\rangle_{NS}$ back to the NS sector.}:
\bea \label{SpFlwN}
|m\rangle_R &=& \cS^{2m+1} |0\rangle_{NS} \,.
\eea
The dual geometries were constructed in \cite{Lunin:2004uu,Giusto:2004id,*Giusto:2004ip}, and they are reviewed in Section \ref{Sec:SpecFlow}. This section also discusses the geometry corresponding to the deformation (\ref{JonPsi}) with 
$|\psi\rangle=|m\rangle_R$. Although equation (\ref{SpFlwN}) completely specifies the state, a more explicit description is known in terms of elements of the superconformal algebra. As discussed in \cite{Giusto:2004id,*Giusto:2004ip}, the state $|m\rangle_R$ is obtained by application of $SU(2)$ generators ${\cal J}^-_{-n}$ to the Ramond vacuum:
\bea
|m\rangle_R &=& 
({\cal J}^-_{-2m})^{n_1n_5}\dots
({\cal J}^-_{-4})^{n_1n_5}
({\cal J}^-_{-2})^{n_1n_5} |0\rangle_{R} \,.
\eea
We shall come back to the general state $|m\rangle_R$ in Section \ref{Sec:SpecFlow}. In the next section however, we take the case of $m=0$ and present a first explicit example of our construction.

\section{Example 1: Deformation of a D1-D5 background} 
\label{sec:MMexample}
\label{SectMM}
In this section we present an explicit example of our construction. We take the background geometry which corresponds to the CFT state $|0\rangle_R$ defined in \eq{eq:0r}. 
The explicit example proceeds by rewriting the linearized perturbation found in \cite{Mathur:2011gz,Mathur:2012tj} in such a way that the solution is valid at nonlinear order. The form of the resulting solution will motivate the general ansatz of the next section.

\subsection{Background and linear perturbation}

In this section we set the D1 and D5 charges to be equal, $Q_1=Q_5=Q$, where
\bea \label{eq:Q}
Q_1 &=& \frac{g^2 \ap^3}{V} n_1 \,, \qquad Q_5 ~=~  \ap n_5 \,, \qquad (2\pi)^4 V ~=~ \mbox{vol}(T^4) \,.  
\eea
The background geometry corresponding to the CFT state $|0\rangle_R$ was constructed in \cite{\MM} and takes the following form\footnote{We use null coordinates $u=t+y$ and $v=t-y$, and use the shorthand notation $c_\theta$, $s_\theta$ for the trigonometric functions $\cos\theta$, $\sin\theta$. Throughout the paper, we raise and lower $T^4$ indices $\a$ with the flat metric $\delta_{\a\beta}$.}:
\bea\label{D1D5MM}
ds^2&=&-\frac{1}{H}\left[du+A\right]\left[dv+B\right]+
Hf\left[\frac{dr^2}{r^2+a^2}+d\theta^2\right]
+H\left[r^2c_\theta^2 d\psi^2+(r^2+a^2)s_\theta^2d\phi^2\right]
\nonumber\\
&& {} \qquad\qquad\qquad\qquad\qquad\qquad\qquad\qquad\qquad\qquad\qquad\qquad\qquad\qquad~~        
+dz^\alpha dz^\alpha \,,\nn
C^{(2)}&=&\frac{1}{2H}[dv+B]\wedge [du+A]+Q \, c_\theta^2 \, d\phi\wedge d\psi \,, 
\eea
where
\be
A=\frac{aQ}{f}\{s_\theta^2 d\phi-c_\theta^2 d\psi\} \,,\quad
B=\frac{aQ}{f}\{s_\theta^2 d\phi+c_\theta^2 d\psi\} \,,\quad
f=r^2+a^2c^2_\theta \,,\quad H=1+\frac{Q}{f} \,. 
\ee
The null coordinates $u$ and $v$ are related to $t$ and $y$ as follows:
\bea\label{DefUV}
u=t+y,\qquad v=t-y,\qquad y~\sim~ y+2\pi R_y,\qquad a=\frac{Q}{R_y} \,.
\eea

Let us single out a direction on the torus and label it by $z$. We now review the linearized perturbation corresponding to the state
\bea \label{eq:J_0r}
J^z_{-n}|0\rangle_R 
\eea
in the limit where the added energy is small. 
The perturbation takes the form~\cite{Mathur:2011gz,Mathur:2012tj}
\be\label{FirstOrderA}
ds^2 = ds^2_0 + \e \, e^{-in\frac{v}{R_y}}
\left(\frac{r^2}{r^2+a^2}\right)^{n/2}\left\{
\frac{Q}{Q+f}\left[
dv-a(c_\theta^2 d\psi+s_\theta^2 d\phi)\right]+
\frac{iaQ}{r(r^2+a^2)}dr
\right\}dz  \qquad 
\ee
\be%
C^{(2)} = C^{(2)}_0 + \e \, e^{-in\frac{v}{R_y}}
\left(\frac{r^2}{r^2+a^2}\right)^{n/2} dz \wedge \left\{
\frac{Q}{Q+f}\left[
dv-a(c_\theta^2 d\psi+s_\theta^2 d\phi)\right]+
\frac{iaQ}{r(r^2+a^2)}dr
\right\}  \qquad \nonumber
\ee
which is a solution to leading order in $\e$.

\subsection{Nonlinear deformation}

The main goal of this section is to promote (\ref{FirstOrderA}) into an exact solution of supergravity equations by making an educated guess and checking it. We begin with rewriting \eq{FirstOrderA} in a form closer to that of \eq{D1D5MM}:
\be\label{FirstOrderA1}
ds^2_1 = ds^2_0 + \e \, e^{-in\frac{v}{R_y}}
\left(\frac{r^2}{r^2+a^2}\right)^{n/2}\left\{-
\frac{f}{Q+f}[dv+B]+
dv+
\frac{iaQ}{r(r^2+a^2)}dr
\right\}dz  \qquad 
\ee
\be%
C^{(2)}_1 = C^{(2)}_0 + \e \, e^{-in\frac{v}{R_y}}
\left(\frac{r^2}{r^2+a^2}\right)^{n/2} dz \wedge \left\{-
\frac{f}{Q+f}[dv+B]+
dv+
\frac{iaQ}{r(r^2+a^2)}dr
\right\}  \qquad \nonumber
\ee
Now we note that the last two terms in the braces above form a complete differential:
\bea
\e \, e^{-in\frac{v}{R_y}}
\left(\frac{r^2}{r^2+a^2}\right)^{n/2}\left[
dv+\frac{ia Q dr}{r(r^2+a^2)}
\right]
&=&
\e \, d\left[\frac{i R_y}{n} e^{-in\frac{v}{R_y}}
\left(\frac{r^2}{r^2+a^2}\right)^{n/2}\right] ~=~
\e \, d\Psi \,, \nonumber
\eea
where the last equality defines the function $\Psi$. As a result we can gauge them away from the metric and $C^{(2)}$ by an infinitesimal diffeomorphism and gauge transformation:
\bea \label{eq:torus_diffeo}
g_{\m\n} &=& (g_1)_{\m\n} + \e \, \xi_{(\mu;\nu)} \,, \qquad\qquad \xi_z ~=~ -\Psi \,, \cr
C^{(2)} &=& C^{(2)}_1 + \e \, d(\Psi dz)  \,.
\eea
We then have the solution
\bea\label{FirstOrderA2}
ds^2 &=& ds^2_0 + \e \, e^{-in\frac{v}{R_y}}
\left(\frac{r^2}{r^2+a^2}\right)^{n/2}\left\{-
\frac{f}{Q+f}[dv+B]\right\}dz \,, \qquad \cr
C^{(2)} &=& C^{(2)}_0 + \e \, e^{-in\frac{v}{R_y}}
\left(\frac{r^2}{r^2+a^2}\right)^{n/2}  
\frac{f}{Q+f}[dv+B] \wedge dz \,.
\,.
\eea
A direct check now reveals that this solution is in fact valid at full nonlinear order with $\e=1$. Thus from now onwards we set $\e=1$. Since we are dealing with a nonlinear solution, we also take the real part of the above fields.

We now observe that the deformed solution (\ref{FirstOrderA2}) can be naturally embedded in the original ansatz \eq{D1D5MM}. We can write the  deformation by adding a new term to field $A$ appearing in the metric and in the two--form:
\bea\label{NewPhoton}
A &\rightarrow& A+ \Phi \, dz \,, \\
\label{PhiMMSp}
\Phi &=& \left(\frac{r^2}{r^2+a^2}\right)^{n/2} \cos{\left(\frac{nv}{R_y}\right)} . 
\eea 
To generalize this solution, we observe that $\Phi$ satisfies the wave equation
\bea\label{NablaPhi}
\nabla^2\Phi(v,x_\mu)&=&0,
\eea 
where $\nabla^2$ is taken with respect to the metric (\ref{D1D5MM}). The general construction presented in the next section guarantees that any $u$--independent solution of
the wave equation (\ref{NablaPhi}) generates a supersymmetric geometry via deformation (\ref{NewPhoton}) of the background (\ref{D1D5MM}). 
The most general $u$--independent solution of the wave equation (\ref{NablaPhi}), which remains finite everywhere, can be written as a superposition of functions (\ref{PhiMMSp}) (along with sine solutions):
\bea\label{PhiFourier}
\Phi=\sum_{n=-\infty}^\infty c_n e^{-in\frac{v}{R_y}}
\left(\frac{r^2}{r^2+a^2}\right)^{|n|/2} \,.
\eea
Since $\Phi$ is real, we have $(c_n)^*=c_{-n}$.
The form of $\Phi$ ensures that the value of function 
$\Phi(v,x_\mu)$  is uniquely determined by the data at infinity, 
\bea\label{fOfV}
f(v)=\lim_{r\rightarrow\infty} \Phi(v,x_\mu)=\sum_{n=-\infty}^\infty c_n.
\eea
Equations (\ref{PhiFourier}) and (\ref{fOfV}) give an implicit map between $f(v)$ and $\Phi(v,x_\mu)$ (one has to expand the periodic function $f(v)$ in the Fourier series and substitute the coefficients in (\ref{PhiFourier})), and in Appendix 
\ref{AppGreen} we derive a more explicit integral transform:
\bea\label{MainGreen}
\Phi=\int_0^{2\pi R_y}\frac{dv'}{2\pi R_y}f(v')
\frac{1-q^2}{
1+q^2-2q\cos\frac{v-v'}{R_y}}.
\eea 
To summarize the results of this section, we found that the solution (\ref{D1D5MM}) can be deformed by making a shift (\ref{NewPhoton}) with $\Phi$ given by (\ref{PhiFourier}). 
We record the whole solution for later convenience:
\bea\label{MMConjIn}
\label{MMConj}
ds^2&=&-\frac{1}{H}\left[du+A\right]\left[dv+B\right]+
Hf\left[\frac{dr^2}{r^2+a^2}+d\theta^2\right]
+H\left[r^2c_\theta^2 d\psi^2+(r^2+a^2)s_\theta^2d\phi^2\right]
\nonumber\\
&& {} \qquad\qquad\qquad\qquad\qquad\qquad\qquad\qquad\qquad\qquad\qquad\qquad\qquad\qquad
+dz^\alpha dz^\alpha \,,\nn
C^{(2)}&=&\frac{1}{2H}[dv+B]\wedge [du+A]+Q \, c_\theta^2 \, d\phi\wedge d\psi \,, 
\eea
\be
A=\frac{aQ}{f}\{s_\theta^2 d\phi-c_\theta^2 d\psi\} + \Phi \, dz_1 \,,\quad
B=\frac{aQ}{f}\{s_\theta^2 d\phi+c_\theta^2 d\psi\} \,,\quad
f=r^2+a^2c^2_\theta \,,\quad H=1+\frac{Q}{f} \,, \nonumber
\ee
\bea \label{PhiMM}
\Phi&=&\sum_{n=-\infty}^\infty c_n e^{-in\frac{v}{R_y}}
\left(\frac{r^2}{r^2+a^2}\right)^{|n|/2} \,.
\eea
The global properties of the deformed geometry (\ref{MMConjIn})--(\ref{PhiMM}) and the map to the CFT state $e^{\mu J^\a_{-n}}|0\rangle_R$ will be discussed in Section \ref{sec:properties}.

\section{Deformation of a general GMR background} \label{sec:GMRetc}

In this section our deformation will be applied to solutions of type IIB string theory obtained by a trivial lifting of a supersymmetric configuration of minimal supergravity in six dimensions. 
The local form of such a background solution is described by the GMR formalism \cite{\GMR}, which is reviewed in Appendix \ref{AppGMR}. 
Since the deformation involves the additional directions, the deformed solutions lie outside minimal six--dimensional supergravity. 

As discussed in the Introduction, we focus on backgrounds which are smooth and horizonless.
While the GMR formalism provides the local structure of solutions, regularity imposes some restrictions on the GMR data, which are not understood completely. 
The regularity conditions are known for the the two-charge D1--D5 geometries \cite{\lmAdS,\lmm} and for some special cases of the three charge system \cite{Lunin:2004uu,Giusto:2004id,*Giusto:2004ip} which will be studied in the next section.
In each of these cases, the deformation generates new smooth and horizonless solutions. We expect that this is a general property of our construction; i.e.~that given a smooth and horizonless background, the deformed geometry is also smooth and horizonless.

We first present the deformation of the general GMR ansatz, then discuss a class of two-charge D1-D5 solutions which lie in this ansatz (in particular, this implies that $Q_1=Q_5$). We also discuss possible generalizations of these results beyond minimal supergravity in six dimensions.

\subsection{The generating technique}
\label{SectSSgrn}

We begin with the set of supersymmetric solutions of type IIB supergravity which are obtained by a trivial lifting of a solution to 6D minimal supergravity. We write such a solution in local GMR form as\footnote{Our conventions are related to those of \cite{Gutowski:2003rg} via $u=\sqrt{2}\,v_{GMR}$, $v=\sqrt{2}\,u_{GMR}$, $\beta=\sqrt{2}\,\beta_{GMR}$, $\omega=\sqrt{2}\,\omega_{GMR}$, and $\hat{u}=+$, $\hat{v}=-$.}:
\bea\label{GMRsln}
ds^2&=&-e^{\hat v}e^{\hat u}+Hds_4^2+
dz^\alpha dz^\alpha \,, \nonumber\\
C^{(2)}&=&\frac{1}{2} e^{\hat v}\wedge e^{\hat u} +
e^{\hat v}\wedge {\cal A}+\sigma_2 \,,\\
e^{\hat v}&=&H^{-1}(dv+\beta) \,,\qquad\quad 
e^{\hat u}=du+\omega+\frac{\cF H}{2}e^{\hat v} \,. \phantom{+\Phi_\alpha(v,x) dz^\alpha} 
\nonumber
\eea
Here $ds_4^2$ is a metric on a four dimensional almost hyperkahler base (which can be $v$--dependent), $H$ and $\cF$ are functions and $\beta$ and $\omega$ are 1-forms on this 4-dimensional space. 
The expressions for ${\cal A}$ and $\sigma_2$, as well as equations satisfied by $H,\cF,\beta,\omega$, are presented in Appendix \ref{AppGMR}. 

Assuming a background solution of the form \eq{GMRsln}, we use the result of the last section as an inspiration and seek a deformation of the following local form:
\bea\label{GMRdfm}
ds^2&=&-e^{\hat v}e^{\hat u}+Hds_4^2+
dz^\alpha dz^\alpha \,, \nonumber\\
C^{(2)}&=&\frac{1}{2} e^{\hat v}\wedge e^{\hat u} +
e^{\hat v}\wedge {\cal A}+\sigma_2 \,,\\
e^{\hat v}&=&H^{-1}(dv+\beta) \,,\qquad\quad 
e^{\hat u}=du+\omega+\frac{\cF H}{2}e^{\hat v}+\Phi_\alpha(v,x) dz^\alpha \,. \nonumber
\eea
An explicit check performed in Appendix \ref{AppDeform} demonstrates that (\ref{GMRdfm}) is a new supersymmetric solution if the functions $\Phi_\alpha$ satisfy the wave equation in ten dimensions. This equation can be rewritten in six--dimensional language, obtaining:
\bea\label{GMRLapl}
d\left(*_6\left[D\Phi_\alpha+\d_v\Phi_\alpha He^{\hat v}\right]\right)&=&0 \,.
\eea
In the above equation the 6D metric and the derivative $D\Phi$ are given by
\bea
ds_6^2&=&-e^{\hat v}e^{\hat u}+Hds_4^2 \,,\qquad\quad
D\Phi ~=~ \tilde{d}\Phi-\beta\,\d_v\Phi \,
\eea
where $\tilde d$ is the exterior derivative on the base space,
\bea \label{eq:dtilde}
\tilde d \Phi &=& \d_i \Phi \, dx^i \,.
\eea
We will also use an alternative form of equation (\ref{GMRLapl}) which involves the Hodge duality with respect to the four--dimensional base space with metric $ds_4^2$:
\bea\label{GMRLapl4}
d([*_{4} D\Phi]\wedge He^{\hat v})&=&0\,.
\eea
To summarize, any solution of the form \eq{GMRsln} can be deformed into a new solution given by \eq{GMRdfm} and \eq{GMRLapl4}.

\subsection{Deformations of D1-D5 solutions} \label{sec:D1D5-def}

The local structure of our deformation of the GMR fields is given in (\ref{GMRdfm}) and (\ref{GMRLapl4}).
Since we are interested in smooth and horizonless solutions,
in this section we focus on a large class of GMR solutions for which the regularity requirements are known and analyze the corresponding deformations. 

\subsubsection{Local form of the deformation}

Let us assume that the various functions entering (\ref{GMRsln}) have no $v$--dependence, that the base metric $ds_4^2$ is flat, and that $\cF=0$. Doing so, one arrives at a class of solutions describing microstates of the D1--D5 system 
\cite{\lmAdS}. These geometries are parameterized by a harmonic function $H$ and two vector functions satisfying the self--duality conditions. We write the solution as a generalization of (\ref{D1D5MM}):
\bea\label{D1D5Gen}
ds^2&=&-\frac{1}{H}\left[du+A\right]\left[dv+B\right]+
Hdx^idx^i+dz^\alpha dz^\alpha,\nonumber\\
C^{(2)}&=&\frac{1}{2H}[dv+B]\wedge [du+A]+
{\cal C}_{ij}dx^idx^j,\\
&&dA=*_4 dA,\quad dB=-*_4 dB,\quad d{\cal C}=-*_4dH \,.
\nonumber
\eea
Although the equations for the fields $A$, $B$ and $H$ decouple, these three ingredients must be related to each other to ensure the regularity of the solution. This relation is also required for making a microscopic interpretation. In \cite{\lmAdS} the geometries corresponding to the microscopic states of the D1--D5 system had been parameterized by a string profile ${\bf F}(w)$\footnote{In contrast to the notation of \cite{\lmAdS,\lmm}, we have used $w$ to denote the integration variable in (\ref{StrProfile}) to avoid a confusion with $v$, which plays an important role in the current paper.}:
\bea\label{StrProfile}
H=1+\frac{Q}{L}\int_0^{L}
\frac{dw}{|{\bf x}-{\bf F}|^2},\quad
A_i+B_i=\frac{Q}{L}\int_0^{L}
\frac{{\dot{F}}_idw}{|{\bf x}-{\bf F}|^2},\quad
|{\dot{\bf F}}|^2=1,\quad L=\frac{2\pi\alpha' n_5}{R_y}.
\eea
and in \cite{\lmm} it was demonstrated that all such solutions were regular, as long as ${\bf F}(w)$ did not have self--intersections\footnote{The discussion of \cite{\lmAdS,\lmm} applies to  more general states which have $|{\dot{\bf F}}|\ne 1$, but they go outside of minimal 6D SUGRA, and here we focus on the GMR ansatz. We comment on more general solutions in Section \ref{SectBW}.}. 

Applying the construction of the previous section, we find the deformation of the solution (\ref{D1D5Gen}):
\bea\label{D1D5Def}
ds^2&=&-\frac{1}{H}\left[du+A+
\Phi_\alpha dz^\alpha\right]\left[dv+B\right]+
Hdx^idx^i+dz^\alpha dz^\alpha,\nonumber\\
C^{(2)}&=&\frac{1}{2H}[dv+B]\wedge 
[du+A+\Phi_\alpha dz^\alpha]+
{\cal C}_{ij}dx^idx^j\\
&&dA=*_4 dA,\quad dB=-*_4 dB,\quad d{\cal C}=-*_4dH,\nonumber\\
\label{D1D5Lapl}
&&\qquad\qquad d([*_{4} D\Phi_\alpha]\wedge [dv+B])=0 \,.
\eea
The local analysis presented in the Appendix \ref{AppDeform} proves that this construction solves the equations for the Killing spinors and field equations.

\subsubsection{Regularity and uniqueness}

We next focus on global properties of the metric (\ref{D1D5Def}). We assume that the undeformed geometry comes from the microscopic construction (\ref{StrProfile}), then, as shown in \cite{\lmm}, solution (\ref{D1D5Gen}) is regular everywhere and asymptotically flat.

One can demonstrate that the deformed solution (\ref{D1D5Def}) has the same properties if  
scalars $\Phi_\alpha$ satisfy the conditions (b)--(c) listed below. In the special case this is proven in Section \ref{sec:MM_glob}, here we just state the general result (the justification is outlined on page \pageref{PageD1D5}):

\begin{quote}
Any regular solution (\ref{D1D5Gen}) of the D1--D5 system can be deformed into a regular solution (\ref{D1D5Def}), as long as each function $\Phi_\alpha$ satisfies three conditions:
\label{PageCndPhi}
\begin{enumerate}[(a)]
	\item $\Phi_\alpha$ solves the differential equation (\ref{D1D5Lapl});
	\item $\Phi_\alpha$ remains finite for all values of ${\vec x}$;
	\item $\Phi_\alpha$ approaches a regular function $f_\alpha(v)$ as one goes to infinity on the base.
\end{enumerate}

\end{quote}
In Appendix \ref{AppUnq} we demonstrate that for every periodic function $f_\alpha(v)$, if the solution satisfying the requirements (a)--(c) exists, then it is unique. The boundary data $f_\alpha(v)$ has a direct connection with the CFT: it can be recovered by looking at the combination of currents appearing in (\ref{JonPsi}), and we will discuss this relation in Section \ref{sec:properties}.   

\subsubsection{Asymptotic behaviour} \label{sec:D1D5-asym}

Finally we discuss the implication of the regularity conditions (\ref{StrProfile}) for the deformed solution.
We begin with reviewing the situation for the undeformed solution (\ref{D1D5Gen}), (\ref{StrProfile}). 
The relation $|{\dot{\bf F}}|^2=1$ determines the overall size of the profile as a function of $R_y$. This is the generalization of the relation $a=\frac{Q}{R_y}$ discussed in the previous section (see (\ref{DefUV})). Here $2\pi R_y$ represents the size of the compact $y$ coordinate at infinity, where the metric becomes flat:
\bea
ds^2&=&-dt^2+dy^2+dx^idx^i+dz^\alpha dz^\alpha.
\nonumber
\eea  
In the deformed case, the relation $|{\dot{\bf F}}|^2=1$ between the size of string profile and the radius $R_y$ of the $y$ coordinate remains unchanged. 

Importantly however, $y$ is no longer the appropriate periodic coordinate at infinity. At large values of $r^2\equiv x^ix^i$ metric (\ref{D1D5Def}) becomes flat,
\bea \label{LeadFlatA-1}
ds^2&=&-[du+f_\alpha dz^\alpha]dv+dx^idx^i+
dz^\alpha dz^\alpha \,.
\eea
While the above metric is flat, it is not written in the standard coordinates.
To put (\ref{LeadFlatA-1}) in the canonical form, we remove 
the off--diagonal terms by the diffeomorphism\footnote{We recall that we raise and lower $T^4$ indices with the flat metric, so here $z_\a=z^\a$.}
\bea
&&z'_\alpha=z_\alpha-\frac{1}{2}\int f_\alpha dv,\quad
u'=\lambda\left[u+\frac{1}{4}\int f_\alpha f_\alpha dv\right],\quad
v'=\frac{v}{\la},\nonumber\\
&&\label{LmbdBoostA}
\la^{-2}=1+\frac{1}{8\pi R_y}\int_0^{2\pi R_y} f_\alpha f_\alpha dv  \,.
\eea
where $z'^\alpha$ has the same period as $z^\alpha$, and the constant $\lambda$ takes the value which ensures that $t'=\ha(u'+v')$ is a single valued coordinate:
\bea
t'(y=2\pi R_y)-t'(y=0)&=&\pi R_y\left[\la-\frac{1}{\la}\right]+\frac{\la}{8}\int_0^{2\pi R_y} f_\alpha f_\alpha dv~=~0 \,.
\nonumber
\eea
We define $2 \pi R$ to be the periodicity of the $y'$ coordinate, where $y'=\ha(u'-v')$, which as we have seen, is the appropriate coordinate at infinity:
\bea
2\pi R & \equiv & y'(y=2\pi R_y)-y'(y=0)~=~\pi R_y\left[\la+\frac{1}{\la}\right]+\frac{\la}{8}\int_0^{2\pi R_y} f_\alpha f_\alpha dv \,.
\eea
The resulting metric is
\bea
ds^2&=&-(dt')^2+(dy')^2+dx^idx^i+dz'^\alpha dz'^\alpha \,.
\nonumber
\eea  
The deformation of a given state is constructed by introducing $\Phi_\alpha$, while keeping $n_1,n_5$ and the asymptotic radius $R$ fixed. 
This implies that the deformed solution has
\bea\label{PerPrm}
R_y=\la R.
\eea
We also introduce another useful notation,
\bea\label{DefHAlpha-1}
h_\alpha(v')\equiv f_\alpha(v)=f_\alpha(\la v') 
\eea
and observe that
\bea
\la^{-2}=1+\frac{1}{8\pi R}\int_0^{2\pi R} h_\alpha h_\alpha dv'.
\eea

To summarize, one can start with solution (\ref{D1D5Gen}) characterized by $n_1,n_5,R$ and the profile ${\bf F}_0(w)$ and introduce four periodic functions $h_\alpha(v')=h_\alpha(v'+2\pi R)$. Then the parameters $R_y$, $f_\alpha$ and ${\bf F}(w)$ corresponding to the regular deformation are given by\footnote{The periodicity of $w$, $w \sim w+L$, comes from \eq{StrProfile}.}
\bea
&&\la^{-2}=1+\frac{1}{8\pi R}\int_0^{2\pi R} h_\alpha h_\alpha dv',\qquad R_y=\la R,\qquad f_\alpha(v)=h_\alpha(\frac{v}{\la}),
\cr
&&{\bf F}(w)=\frac{1}{\la}{\bf F}_0(\l w) \,, \qquad w ~\sim~ w + \frac{2 \pi \ap n_5}{R_y} \,.
\eea
Then functions $\Phi_\alpha$ satisfying conditions (a)--(c) in page \pageref{PageCndPhi} lead to the unique regular deformation (\ref{D1D5Def}), which has the same asymptotics as the original solution. 

\subsection{Generalization beyond minimal supergravity}
\label{SectBW}

The deformation presented in this section can be applied to solutions of IIB supergravity, which cannot be lifted from the minimal SUGRA in six dimensions. Such geometries are expected to be important for understanding physics of black holes, since even in the D1--D5 case most microscopic states correspond to profiles which do not satisfy the condition 
$|{\dot{\bf F}}(w)|=1$. In this subsection we will propose 
the deformation of the solutions constructed in \cite{\CarMcCon,\bgsw} and outline the applications of such deformation, if it exists. This class contains, as special cases, all the microstate geometries of the D1--D5 system studied in \cite{\lmAdS}.

Upon lifting to ten dimensions, solutions of \cite{\CarMcCon,\bgsw} produce the following geometry\footnote{We are writing all metrics in the string frame. The dilaton is denoted by $\varphi$.} 
\bea\label{BGSWsoln}
ds^2&=&-e^{\hat v}e^{\hat u}+Hds_4^2+
e^{\varphi}dz^\alpha dz^\alpha\nonumber\\
C^{(2)}&=&\frac{1}{2}e^{-\varphi} e^{\hat v}\wedge e^{\hat u} +
e^{\hat v}\wedge {\cal A}+\sigma_2,\\
e^{\hat v}&=&H^{-1}(dv+\beta),\quad 
e^{\hat u}=du+\omega+\frac{\cF H}{2}e^{\hat v} \,. \nonumber
\eea
Supersymmetry conditions lead to equations for $H$, $\cF$, $e^{\varphi}$, $\beta$, $\omega$, ${\cal A}$, $\sigma_2$, which have a structure very similar to the one reviewed in the Appendix \ref{AppGMR}. We refer to \cite{\CarMcCon,\bgsw} for the explicit construction. We conjecture that any solution (\ref{BGSWsoln}) can be deformed by making a replacement
\bea\label{BGSWsoln-2}
e^{\hat u}\rightarrow du+\omega+\frac{\cF H}{2}e^{\hat v}+
\Phi_\alpha dz^\alpha,
\eea
where $\Phi_\alpha$ satisfy (\ref{GMRLapl4}). To verify this conjecture, one would have to apply the construction presented in the Appendix \ref{AppDeform} to solutions of  \cite{\CarMcCon,\bgsw}, and we leave this for future work. 

In particular, if deformation (\ref{BGSWsoln-2}) indeed generates a new supersymmetric geometry, it can be applied to the general D1-D5 solutions analyzed in \cite{\lmAdS,\lmm} to produce a counterpart of (\ref{D1D5Def}):
\bea\label{GnD1D5Def}
ds^2&=&-\frac{1}{H}\left[du+A+
\Phi_\alpha dz^\alpha\right]\left[dv+B\right]+
Hdx^idx^i+e^{\varphi}d{z}_\alpha dz^\alpha,\nonumber\\
C^{(2)}&=&\frac{e^{-\varphi}}{2H}[dv+B]\wedge 
[du+A+\Phi_\alpha dz^\alpha]+
{\cal C}_{ij}dx^idx^j\\
&&H=\sqrt{f_1f_5},\quad e^{2\varphi}=\frac{f_1}{f_5},\quad
\Delta_4 f_1=\Delta_4 f_5=0\nonumber\\
&&dA=*_4 dA,\quad dB=-*_4 dB,\quad d{\cal C}=-*_4df_5,\nonumber\\
\label{D1D5Lapl-2}
&&\qquad\qquad d([*_{4} D\Phi_\alpha]\wedge [dv+B])=0
\eea
As demonstrated in \cite{\lmAdS,\lmm}, microscopic geometries with $\Phi_\alpha=0$ are parameterized in terms of the string profile  ${\bf F}(w)$:
\bea
f_1=1+\frac{Q_5}{L}\int_0^{L}
\frac{|{\dot{\bf F}}|^2dw}{|{\bf x}-{\bf F}|^2},\quad
f_5=1+\frac{Q_5}{L}\int_0^{L}
\frac{dw}{|{\bf x}-{\bf F}|^2},\quad 
A_i+B_i=\frac{Q_5}{L}\int_0^{L}
\frac{{\dot{F}}_idw}{|{\bf x}-{\bf F}|^2},\nonumber
\eea
and they remain regular everywhere. The deformed geometry, if it exists, 
is also regular as long as $\Phi_\alpha$ satisfy the requirements (a)--(c) in page \pageref{PageCndPhi}.

The solutions (\ref{GnD1D5Def}) correspond to excitations of the six--dimensional space, and excitations of the torus have also been constructed in \cite{\lmm,\kst}. 
It is natural to conjecture that our generating technique is applicable to these solutions as well; we leave a full investigation for future work. The conjecture is that a new solution is obtained  by making the replacements
\bea
dt\rightarrow dt+\frac{1}{2}\Phi_\alpha dz^\alpha,\qquad
dy\rightarrow dy+\frac{1}{2}\Phi_\alpha dz^\alpha
\eea
in equation (B.2) of \cite{\lmm} or in equation (2.11) of \cite{\kst}. We expect that this deformation will preserve the regularity of the solutions, as in the cases studied in the present paper.

\section{Example 2: Deformation of a class of D1-D5-P backgrounds}
\label{Sec:SpecFlow}

In this section we apply the $J^\a_{-n}$ deformation to the class of three-charge D1-D5-P supersymmetric solutions found in \cite{Lunin:2004uu,Giusto:2004id,*Giusto:2004ip}. The corresponding CFT states were reviewed in Section \ref{SectCFT} and denoted $|m\rangle_R$ where $m$ is an integer parameter controlling the amount of spectral flow from the NS vacuum.

The solutions of  \cite{Lunin:2004uu,Giusto:2004id,*Giusto:2004ip} take the standard GMR form lifted to 10D \cite{Giusto:2004kj},
\bea
ds^2 &=& -H^{-1}(dv +  \beta)\Bigl(du +  \omega + {\cF\over 2}(dv +  \beta)\Bigr) + H ds^2_4 + dz^\alpha dz^\alpha \,,
\label{general6d}
\eea
where $H$, $\cF$, $\beta$ and $\o$ are given by (we again use $c_{\theta}=\cos\th, s_\theta=\sin\th$)
\bea
H&=&1+\frac{Q}{f} \,, \qquad\quad
{\cF\over 2 } ~=~ -{Q_p\over f} \,, \qquad\quad
\beta~=~ {Q\over f}\,a \,\eta\,(c_{\theta}^2 d\psi
+ s_\theta^2 d\phi) \,, \nonumber\\
\omega&=& {Q\over f} \Bigl[\Bigl(2\g_1 -
a  \,\eta\,\Bigl(1-2{Q_p\over f}\Bigr)\Bigr)c_{\theta}^2 d\psi 
+ \Bigl(2\g_2 - a \,\eta\,\Bigl(1-2{Q_p\over f}\Bigr)\Bigr)s_\theta^2 d\phi\Bigr] 
\label{3chargegmr}
\eea
and where
\bea
f &=& r^2+ a \,\eta\,\bigl(\g_1 \, s_\theta^2 + \g_2 \,c_{\theta}^2 \bigr) \,, 
\qquad Q_p ~=~ -\g_1\g_2\,,\qquad \eta ~=~ {Q\over Q + 2Q_p} \,, \nonumber\\
\g_1 &=& - a\,m\,,\quad \g_2=a \big( m+1 \big),   \quad m\in\mathbb{Z} \,. 
\eea
Here $Q=Q_1=Q_5$ is given in terms of $n_1$ and $n_5$ by \eq{eq:Q}.
The base metric has the following form: 
\bea
ds^2_4 &=& f\Bigl({dr^2\over r^2+a^2\, \eta} + d\theta^2\Bigr)\nonumber\\
&& {} + {1\over f}\Bigl[[r^4 +
r^2\,a\,\eta\,(2\g_1 + a (2m+1) \, c_{\theta}^2) +
a^2\,\g_1^2\,\eta^2\,s_\theta^2]c_{\theta}^2\,d\psi^2 -2\g_1 \g_2
\,a^2\, \eta^2\,s_\theta^2 c_{\theta}^2 \,d\psi d\phi \quad 
\nonumber\\
&& {} \qquad +[r^4 + r^2\,a\,\eta\,(2\g_2 - a (2m+1) \,s_\theta^2) +
a^2 \,\g_2^2 \,\eta^2 \,c_{\theta}^2]s_\theta^2
\,d\phi^2
\Bigr] \,.
\label{3chargebase}
\eea
We omit the background $C^{(2)}$ here, but we shall write the deformed $C^{(2)}$ shortly.
As in Example~1 studied in Section \ref{sec:MMexample}, the null coordinates $u$ and $v$ are related to $t$ and $y$ as follows:
\bea\label{MMSpFlaRy}
u=t+y,\qquad v=t-y,\qquad y~\sim~ y+2\pi R_y,\qquad a=\frac{Q}{R_y} \,.
\eea
The last condition ensures that the metric is everywhere regular and has no horizon. It will be useful to think of this condition as defining $a$ in terms of $n_1, n_5, R_y$.

We now apply the general results of Section \ref{SectSSgrn} to find the $J^z_{-n}$ deformation on these states, 
\bea \label{eq:J_0405_state}
J^z_{-n} |m \rangle_R \,.
\eea
To solve the wave equation on these backgrounds, we make an ansatz 
\bea
\Phi &=& f(r) \cos{\left(\frac{nv}{R_y}\right)} \,
\eea
and find the solution
\bea
\Phi &=& \left( \frac{r^2}{r^2[1+2a^2 m (m+1)] + a^2 }  \right)^{\frac{n}{2}} \cos{\left(\frac{nv}{R_y}\right)} \,.
\eea
Note that for $m=0$ this solution reduces to the deformation studied in Example 1, c.f.~\eq{PhiMMSp}.
We have explicitly checked the field equations with the above deformation.

To make the discussion more general, we introduce four deformations $\Phi_\alpha$:
\bea\label{PhiAlpha-SF}
\Phi_\alpha &=& \sum_n c^\alpha_n\left( \frac{r^2}{r^2[1+2a^2 m (m+1)] + a^2}  \right)^{|{n}/{2}|} e^{i{nv}/{R_y}}~ \,
\eea
and for later convenience, we write the full deformed solution. The metric is given by
\bea\label{DfrSpFl}
ds^2 & = & -\frac{1}{H} \bigl[du + \Phi_\a dz^\a \bigr] \, dv + \frac{Q_{p}}{Hf}\, dv^{2}
+ H f \left( \frac{dr^2}{r^2 +a^2\eta} + d\theta^2 \right)\nonumber \\
&& {} + H \Bigl( r^2 + \g_1\,a\,\eta - \frac{Q^2\,(\g_1^2-\g_2^2)\,\eta\,c_\theta^2}{H^2 f^2}\Bigr)
c_\theta^2 d\psi^2  \nonumber \\
&& {} + H\Bigl( r^2 + \g_2\,a\,\eta + \frac{Q^2\,(\g_1^2-\g_2^2)\,\eta\,s_\theta^2}{H^{2} f^{2}}\Bigr) 
s_\theta^2 d\phi^2  \nonumber \\
&& {} + \frac{Q_p\,a^2\,\eta^2}{H f} \left( c_\theta^2 d\psi + s_\theta^2 d\phi \right)^{2} 
 - \frac{2 Q }{Hf}
\left(\g_1 c_\theta^2 d\psi + \g_2 s_\theta^2 d\phi\right)dv \quad 
\nonumber \\
&& {} 
- \frac{a\eta \, Q}{H f}
\left( c_\theta^2 d\psi + s_\theta^2 d\phi \right) \bigl( \bigl[ du + \Phi_\a dz^\a \bigr]-dv \bigr) \,
\eea
and the deformed $C^{(2)}$ is
\bea\label{DfrSpFl-C2} 
C^{(2)} & = & -\frac{1}{2H} \bigl[du + \Phi_\a dz^\a \bigr] \wedge dv 
+ \frac{1}{Hf} \left( a \eta \, Q_p - \frac{aQ}{2} \right) \bigl[du + \Phi_\a dz^\a \bigr] \wedge \left( c_\theta^2 d\psi + s_\theta^2 d \phi \right)
\nonumber \\
&& {} - \frac{Q}{2Hf} a (2m+1) \, dv \wedge \left( c_\theta^2 d\psi - s_\theta^2 d \phi \right)
\nonumber\\
&& {} 
- \frac{Q}{Hf} c_\theta^2 \Bigl( r^2 + a^2 \eta (m+1) + Q \Bigr) d\psi \wedge d \phi \,.
\eea
As in the case of the solutions studied in Section \ref{sec:D1D5-def}, the deformed solution has flat asymptotics, but it does not reduce to the canonical form of the Minkowski metric at infinity. In the next section we shall find the standard coordinates and use them to read off the charges of the solution and identify the corresponding CFT state.

\section{Global properties of the deformed examples}
\label{sec:properties}

While constructing the deformations in Sections \ref{sec:MMexample}, \ref{sec:GMRetc} and \ref{Sec:SpecFlow}, we focused on local analysis. In this section we will address the regularity of the solutions (\ref{MMConjIn})--(\ref{PhiMM}) and \eq{PhiAlpha-SF}--\eq{DfrSpFl-C2} and discuss the map to the dual CFT.  

\subsection{Recovery of standard asymptotics} 
\label{sec:SF-fix-Ry}

To find the map between the new geometries and the states in the CFT, we evaluate the charges of the deformed solutions. We will focus on deformation of D1--D5--P geometries discussed in Section \ref{Sec:SpecFlow}, and the deformation of the Ramond vacuum (\ref{MMConjIn})--(\ref{PhiMM}) can be obtained by setting $m=0$. 
In this subsection we will assume that $c^\alpha_0=0$ in 
(\ref{PhiAlpha-SF}) since the constant terms in $\Phi$ can be removed by shifting $u$, and thus they do not contribute to the charges\footnote{Since $y$ and $z^\alpha$ are periodic directions, such shifts are not genuine diffeomorphisms, and they change {\it global} properties of the solution.}.

At infinity the metric (\ref{DfrSpFl}) approaches flat space:
\bea\label{LeadFlatA}
ds^2&=&-\left[du+f_\alpha(v) dz^\alpha\right]dv+dr^2+r^2d\Omega_3^2+
dz^\alpha dz^\alpha,
\eea
where
\bea\label{FOfVFlowA}
f_\alpha(v)\equiv\lim_{r\rightarrow\infty}\Phi_\alpha(r,v)=
\sum_n c^\alpha_n e^{i{nv}/{R_y}}.
\eea
We now repeat the analysis of asymptotics performed in Section \ref{sec:D1D5-asym}.
The unique diffeomorphism which puts the metric (\ref{LeadFlatA}) in the standard form while keeping $t'=\ha(u'+v')$ a single-valued coordinate is
\bea\label{FOfVFlow-2}
&&z'_\alpha=z_\alpha-\frac{1}{2}\int f_\alpha dv,\quad
u'=\lambda\left[u+\frac{1}{4}\int f_\alpha f_\alpha dv\right],\quad
v'=\frac{v}{\la},\\
&&\label{LmbdBoost}
\la^{-2}=1+\frac{1}{8\pi R_y}\int_0^{2\pi R_y} f_\alpha f_\alpha dv \,.
\eea
In the new coordinates, the metric (\ref{LeadFlatA}) can be written as 
\bea
ds^2&=&-(dt')^2+(dy')^2+dr^2+r^2d\Omega_3^2+
dz'^\alpha dz'^\alpha,
\eea
where $z'^\alpha$ has the same period as $z^\alpha$, and as before 
$y'=\ha(u'-v')$ is defined to have period $2\pi R$:
\bea\label{PeriodPrm}
y'\sim y'+2\pi R,\qquad R_y=\la R \,.
\eea
The deformation of a given state is constructed by introducing $\Phi_\alpha$, while keeping $n_1,n_5,m$ and $R$ fixed.
We also introduce
\bea\label{DefHAlpha}
h_\alpha(v')\equiv f_\alpha(v)=f_\alpha(\la v') 
\eea
and observe that
\bea
\la^{-2}=1+\frac{1}{8\pi R}\int_0^{2\pi R} h_\alpha h_\alpha dv'.
\eea

\subsection{Charges of the solutions and map to CFT}
\label{sec:MM_cft_map}

The derivation of charges is now straightforward. The details are presented in Appendix \ref{app:Charges}, and here we just 
quote the result from (\ref{sumChrgAp}):
\bea\label{sumChrg}
&&J_\phi=\frac{n_1n_5}{2}\left[m+1\right],\qquad
J_\psi=-\frac{n_1n_5}{2}m \,,
\qquad P_\alpha=0 \,,\nonumber\\
&&P_{y'}=\frac{n_1n_5}{R}\left[m(m+1)+\frac{Q}{8\pi R a^2}\int_0^{2\pi R} dy' {h_\alpha h_\alpha}\right] ~=~ P_{y'}^{(0)}+P_{y'}^{(1)}\,. \qquad 
\eea
In the last line we have separated the background $y'$ momentum charge $P_{y'}^{(0)}$ and the momentum added by the perturbation,
\bea
P_{y'}^{(1)} &=& \frac{n_1n_5}{R}
\frac{Q}{8\pi R a^2}\int_0^{2\pi R} dy' {h_\alpha h_\alpha}
\, .
\eea
The above charges are derived from the asymptotically flat region of the geometry and are valid for all values of the parameters $a$ and $Q$. 

In the regime of parameters $a\ll \sqrt{Q}$, the background geometry has a large AdS throat; one can decouple the AdS region from the flat asymptotics and study AdS/CFT in the resulting asymptotically AdS geometry. 
In Appendix \ref{app:Charges} we demonstrate that the same charges (\ref{sumChrg}) can be recovered from the AdS region.

We now use the charges (\ref{sumChrg}) 
as well as expansion (\ref{FOfVFlowA}) 
to map the deformed geometries into the states in the dual CFT. 
To do so, we rewrite 
the last relation in terms of the Fourier coefficients of 
functions $h_\alpha$ (see equations (\ref{FOfVFlowA}), (\ref{DefHAlpha})):
\bea\label{MomShift}
P_{y'}^{(1)} &=& \frac{n_1n_5}{R}
\frac{Q}{2 a^2}\sum_{n>0} c^\alpha_n c^\alpha_{-n}=
 \frac{n_1n_5}{R}
\frac{Q}{2 a^2}\sum_{n>0} {c^\alpha_n}  (c^\alpha_{n})^*
\, .
\eea
This expression should be compared with momentum of the CFT state 
(\ref{JonPsi}), which can be evaluated using the commutation relations
\bea
[L_m,J^\alpha_n]=-nJ^\alpha_{n+m},\quad 
[J^\alpha_m,J^\beta_n]=m
\frac{n_1n_5}{2}\delta^{\alpha\beta}\delta_{m+n}\,.
\eea
Using commutator of the currents, we find the normalized version of the state (\ref{JonPsi}) (assuming that $\langle\psi|\psi\rangle=1$):
\bea\label{JonPsiPrm}
|\Psi\rangle=
\exp\left[-\frac{n_1n_5}{4}\sum_{n>0} n{\bar \mu}^\alpha_n\mu^\alpha_n\right] 
\exp{\left(\sum_{n>0} \mu^\alpha_n J^{\a}_{-n} \right)}|\psi\rangle \,.
\eea
To find the momentum, we evaluate the expectation value of $L_0$:
\bea
\langle\Psi|L_0|\Psi\rangle&=&
\langle\psi|L_0|\psi\rangle+
\sum_{n,\alpha}\left[\exp\left[-\frac{n_1n_5}{2}n{\bar \mu}^\alpha_n\mu^\alpha_n\right]
\left(\sum_{k=0}^\infty\frac{1}{k!}
(nk)(\mu_n^\alpha {\bar\mu}_n^\alpha)^k
\left(\frac{nn_1n_5}{2}\right)^k
\right)\right]\nonumber\\
&=&\langle\psi|L_0|\psi\rangle+
\sum_{n,\alpha}\left[\exp\left[-\frac{n_1n_5}{2}n{\bar \mu}^\alpha_n\mu^\alpha_n\right]
\left(n\mu_n^\alpha
\frac{\d}{\d\mu_n^\alpha}\exp\left[\frac{nn_1n_5}{2} {\bar \mu}^\alpha_n\mu^\alpha_n
\right]\right)\right]\nonumber\\
&=&\langle\psi|L_0|\psi\rangle+
\sum_{n,\alpha}\frac{n^2n_1n_5}{2} {\bar \mu}^\alpha_n\mu^\alpha_n
\eea
Since the right--moving sector is not excited, the expectation value of 
${\bar L}_0$ is not modified, and we find
\bea
\langle\Psi|L_0-{\bar L}_0|\Psi\rangle
&=&\langle\psi|L_0-{\bar L}_0|\psi\rangle+
\sum_{n,\alpha}\frac{n^2n_1n_5}{2} {\bar \mu}^\alpha_n\mu^\alpha_n \,.
\eea
Comparing with (\ref{MomShift}), we find the map
\bea
\mu_n^\alpha \quad \longleftrightarrow \quad \frac{1}{n}\sqrt{\frac{Q}{a^2}}c^\alpha_n \,.
\eea
This provides the relation between the gravity solution and the corresponding CFT state.

\subsection{Deformation of a special state: global geometry} \label{sec:MM_glob}

In this section we demonstrate the regularity of the geometry (\ref{MMConjIn})--(\ref{PhiMM}).
To do so, we rewrite the metric in terms of the $\Phi$--independent part (\ref{D1D5MM}), which will be called $ds_0^2$, and the $\Phi$--dependent contribution:
\bea\label{PhiMetr0}
ds^2&=&ds_0^2
-\frac{1}{H}\Phi dz
\left[dv+\frac{aQ}{f}\{s_\theta^2 d\phi+c_\theta^2 d\psi\}\right] .
\eea
As demonstrated in \cite{\mm}, the original metric $ds_0^2$ remains regular everywhere, as long as $a$ is given by 
\bea\label{MMaRy}
a=\frac{Q}{R_y},
\eea
and we now demonstrate that the deformation preserves this property. 
The determinant of the metric (\ref{MMConjIn}) is  
\bea
\mbox{det} ~g=-\frac{1}{4}(Hf)^2r^2 s_\theta^2 c_\theta^2,
\eea 
so as long as $\Phi$ remains finite, the metric can only become singular at $r=0$, at $\theta=0,\frac{\pi}{2}$, or at $(r,\theta)=(0,\frac{\pi}{2})$, where $f$ goes to zero. Let us analyze the vicinity of these potentially dangerous points.
\begin{enumerate}[(a)]
\item{In the vicinity of the points where $\theta=0$ while $r>0$, coordinate $\phi$ becomes ill-defined, but metric $ds_0^2$ is regular in the coordinates 
\bea\label{Jun25Phi}
r,\ u,\ v,\ \psi,\ x_1=\sin\theta\cos\phi,\ x_2=\sin\theta\sin\phi
\eea
It is clear that the $\Phi$--dependent term in (\ref{PhiMetr0}) is also regular in these coordinates.}
\item{The vicinity of the points $\theta=0$, $r>0$ works in the same way.}
\item{In the vicinity of the points where $r=0$ while 
$\theta\ne \frac{\pi}{2}$, functions $f$ and $H$ remain finite, and the unperturbed metric can be rewritten as
\bea
ds_0^2&\approx&-\frac{1}{H}(dt+\frac{Qa}{f}s_\theta^2 d\phi)^2+Hf d\theta^2+Ha^2 s_\theta^2d\phi^2
+dz^\alpha dz^\alpha\nonumber\\
&&+H\left[r^2+\frac{a^2Q^2 c_\theta^2}{H^2f^2}\right]c_\theta^2
\left(d\psi-\frac{a}{Q}dy\right)^2+2\frac{a}{Q}Hr^2\left[1-\frac{Q^2}{(Hf)^2}\right]c_\theta^2 \left(d\psi-\frac{a}{Q}dy\right)dy
\nonumber\\
&&+\frac{Hf dr^2}{a^2}\left[dr^2+r^2\frac{a^2dy^2}{Q^2}\right].
\nonumber
\eea
The coordinate $y$ becomes ill--defined at $r=0$, but the metric is regular in coordinates
\bea\label{Jun25Psi}
t,\ \theta,\ \phi,\ \psi-\frac{a}{Q}y,\
x_1=r\cos\frac{y}{R_y},\ x_2=r\sin\frac{y}{R_y}.
\eea
It is clear that the deformed metric (\ref{PhiMetr0}) remains regular in coordinates (\ref{Jun25Psi}) as well. 
}
\item{In the vicinity of the ring $(r,\theta)=(0,\frac{\pi}{2})$, inequality $f\ll Q$ is satisfied, then $H\approx Q/f$ and the unperturbed metric can be rewritten as 
\bea
ds^2&\approx&\frac{1}{Q}\left[-(r^2+a^2)dt^2+r^2dy^2\right]+
Q\frac{dr^2}{r^2+a^2}+Q\left[d\theta^2+
s_\theta^2d{\tilde\phi}^2+
c_\theta^2d{\tilde\psi}^2\right]+
dz^\alpha dz^\alpha\,,\nonumber\\
&&{\tilde\phi}\equiv \phi-\frac{t}{R_y},\quad
{\tilde\psi}\equiv \psi-\frac{y}{R_y}\,.\nonumber
\eea
 This metric is regular in the coordinates
\bea\label{NHExpn}
t,\ \phi,\ 
\begin{array}{l}
x_1=r\cos\frac{y}{R_y},\\ 
x_2=r\sin\frac{y}{R_y},
\end{array}\
\begin{array}{l}
x_3=c_\theta \cos\left[\psi-\frac{y}{R_y}\right],\\
x_4=c_\theta \sin\left[\psi-\frac{y}{R_y}\right]
\end{array}\eea
We used the relation (\ref{MMaRy}). Function $\Phi$ given by (\ref{PhiFourier}) remains regular in this coordinates, and equation (\ref{PhiMetr0}) can be rewritten  as
\bea
ds^2&\approx&ds_0^2
-\Phi dz
\left[\frac{r^2}{Q}(dt-dy)+as_\theta^2 d\phi+ac_\theta^2d\left(\psi-\frac{y}{R_y}\right)\right]
\eea
It is clear that the $\Phi$--dependent correction remains regular in coordinates (\ref{NHExpn}). 
}
\end{enumerate}
To summarize, we demonstrated that geometry (\ref{MMConjIn}) is regular everywhere, as long as each of the 
four functions $\Phi$ is given by the expansion (\ref{PhiMM}). Although we only analyzed the metric, one can check that the RR two--form also remains regular in the coordinate systems 
(\ref{Jun25Phi})--(\ref{NHExpn}).

\label{PageD1D5}
We will conclude this subsection by outlining the argument for regularity of the deformation (\ref{D1D5Def}), (\ref{D1D5Lapl}) with $\Phi$ satisfying conditions (a)--(c) in page \pageref{PageCndPhi}. First we observe that in Cartesian coordinates used in (\ref{D1D5Def}), the metric can only become singular when $H$ diverges. In \cite{\lmm}, all undeformed D1--D5 geometries 
(\ref{D1D5Gen})--(\re{StrProfile}) were shown to be regular by utilizing a generalization of the coordinate system (\ref{NHExpn}). These coordinates transform the vicinity of the singular curve, where $H^{-1}=0$, into the standard KK monopole, and they also guarantee regularity of the deformed solution, as long as function $\Phi$ satisfies conditions (a)--(c) in page \pageref{PageCndPhi}.

\subsection{Regularity of the deformed D1--D5--P solution}

Following the logic of the last subsection, one can demonstrate that the deformation of the D1--D5--P solution 
\eq{PhiAlpha-SF}--\eq{DfrSpFl-C2} is also regular. The analysis is slightly more involved since $f$ can now take negative values, and it vanishes on surfaces rather than curves on the base space. We will demonstrate regularity for the special case $a\ll\sqrt{Q}$, which arises in the context of the AdS/CFT correspondence. 

At $r\gg a$, the metric on the base becomes flat and $f\approx r^2$, then the metric 
(\ref{DfrSpFl}) can be approximated by the naive D1--D5--P geometry, which is regular 
away from $r=0$. In the region where $r$ is comparable to (or smaller than) $a$, the assumption $a\ll\sqrt{Q}$ implies that $r\ll\sqrt{Q}$, and function $H$ simplifies:
\bea
H=1+\frac{Q}{f}\approx \frac{Q}{f},\quad 
\eta=\frac{Q}{Q+2a^2m(m+1)}\approx 1 \,.
\eea
With these approximations, the metric (\ref{DfrSpFl}) becomes
\bea\label{AdsFlDef}
ds^2&\approx&\frac{1}{Q}\left[-(r^2+a^2)dt^2+r^2dy^2\right]+
Q\frac{dr^2}{r^2+a^2}+Q\left[d\theta^2+
s_\theta^2d{\tilde\phi}^2+
c_\theta^2d{\tilde\psi}^2\right]+
dz^\alpha dz^\alpha\nonumber\\
&&+\frac{1}{Q}\left[-(r^2+a^2)dt+r^2dy\right]\Phi_\alpha dz^\alpha-a\left[s_\theta^2d{\tilde \phi}+
c_\theta^2 d{\tilde\psi}\right]\Phi_\alpha dz^\alpha \,.
\eea
Here 
\bea
{\tilde\phi}\equiv \phi-(m+1)\frac{t}{R_y}+m
\frac{y}{R_y}=\phi-(m+1)\frac{t}{R_y}+m
\frac{y}{R_y},\quad
{\tilde\psi}\equiv \psi-(m+1)\frac{y}{R_y}+m
\frac{t}{R_y}
\eea
are angles which are globally well-defined. 

It is clear that (\ref{AdsFlDef}) describes a deformation of $AdS_3\times S^3\times T^4$, which is regular everywhere.  This proves regularity of the deformed solution \eq{PhiAlpha-SF}--\eq{DfrSpFl-C2} in the case $a\ll\sqrt{Q}$, and the general case can be analyzed following the last subsection.

\section{Relation to other constructions} \label{sec:pp-wave-gv}

\subsection{Relation to the pp--wave}

Using string dualities, the deformed geometry (\ref{GMRdfm}) can be rewritten in a form where the deformation appears as a pp-wave added to the rotating D1--D5--P system. Let us briefly outline this construction.

We begin with rewriting (\ref{GMRdfm}) in terms of undeformed frames $e_0^{\hat v}$ and $e_0^{\hat u}$:
\bea\label{GMRaltPP}
ds^2&=&-e_0^{\hat v}e_0^{\hat u}+Hds_4^2+
(dz^\alpha -\frac{1}{2}\Phi_\alpha e_0^{\hat v})^2-
\frac{1}{4}\Phi_\alpha \Phi_\alpha (e_0^{\hat v})^2 \,, \nonumber\\
C^{(2)}&=&\frac{1}{2} e_0^{\hat v}\wedge e_0^{\hat u}+
\frac{1}{2} e_0^{\hat v}\wedge \Phi_\alpha dz^\alpha +
e_0^{\hat v}\wedge {\cal A}+\sigma_2 \,,\\
e_0^{\hat v}&=&H^{-1}(dv+\beta) \,,\qquad\quad 
e_0^{\hat u}=du+\omega+\frac{\cF H}{2}e^{\hat v}_0 \,. \phantom{+\Phi_\alpha(v,x) dz^\alpha} 
\nonumber
\eea
The off--diagonal terms containing $dz^\alpha$ can be removed from the metric by performing four T dualities along $z$ directions:
\bea\label{GMRaltPPa}
ds^2&=&-e_0^{\hat v}e_0^{\hat u}-
\frac{1}{4}\Phi_\alpha \Phi_\alpha 
(e_0^{\hat v})^2+Hds_4^2+
dz^\alpha dz^\alpha \,, \nonumber\\
B&=&-\frac{1}{2}\Phi_\alpha e_0^{\hat v}\wedge dz^\alpha,
\qquad
C^{(4)}=
\frac{1}{24} \left[e_0^{\hat v}\wedge \eps_{\alpha\beta\gamma\delta}\Phi_\alpha dz^\beta \wedge
dz^\gamma \wedge dz^\delta+dual\right]\,,
\nonumber\\
C^{(6)}&=&\left[\frac{1}{2} e_0^{\hat v}\wedge e_0^{\hat u} +
e_0^{\hat v}\wedge {\cal A}+\sigma_2\right]dz_1 \wedge dz_2 \wedge dz_3 \wedge dz_4 \,.
\eea
In this duality frame, the deformation $\Phi_\alpha$ can be interpreted as a pp--wave added to the D1--D5--P 
geometry\footnote{Here the D1-D5 system is described by the six--form potential, which is related to the traditional two--form by standard electric--magnetic duality.}. Indeed, upon setting the D1--D5--P charges to zero (i.e., setting $e_0^{\hat v}=dv$, $e_0^{\hat u}=du$, $H=1$),
one gets a combination of the RR and NS--NS pp--waves (see e.g.~\cite{Hull:1984vh,*Blau:2001ne,*Gauntlett:2002cs}). Equation (\ref{GMRaltPPa}) describes the addition of the pp-wave to the rotating D1--D5--P geometry, and it would be interesting to find a procedure for combining pp-waves with other brane configurations by generalizing this construction. 

The above perturbation raises an interesting issue when we discuss the dual CFT. Before we add any perturbations involving the $T^4$, the D1--D5 background was invariant under the operation of applying four T-dualities along the torus directions. But as we have seen above, a perturbation involving the metric component $g_{z\mu}$ changes under these dualities to a perturbation involving $B_{z\mu}$. Restricting our attention to the asymptotically AdS region, we therefore get {\it two} perturbations with similar properties: one from $g_{z\mu}$ and one from $B_{z\mu}$. In the context of the work of Brown and Henneaux \cite{Brown:1986nw}, both these perturbations correspond to asymptotic symmetries. If these symmetries give rise to conserved currents in the CFT, then there should be two such currents.

We recall that in AdS/CFT duality there is a separation of degrees of freedom into those corresponding to the AdS region and the center of mass degrees of freedom which are localized at the boundary of AdS. The asymptotic symmetries of Brown and Henneaux describe degrees of freedom of the latter kind. What we have seen is that this center of mass dynamics appears to have two conserved currents on the same footing. It would be interesting to study this further.

\subsection{Relation to Garfinkle-Vachaspati transform}

In this section we note a similarity of our deformation with the Garfinkle-Vachaspati transform~\cite{Garfinkle:1990jq}.
The Garfinkle-Vachaspati transform is a deformation defined in terms of a vector $k^{\mu}$ which is Killing, null and hypersurface-orthogonal. If there are also additional matter fields present, there are also other conditions.
Given a background metric ${\bar g}_{\m\n}$, the deformation is defined in terms of a harmonic function $\Phi$ which satisfies
$k^\mu \Phi_{;\mu} = 0$ via
\bea
g_{\m\n} &=& {\bar g}_{\m\n} + \Phi \, k_\mu k_\nu \,.
\eea
We observe that our deformation of the general uplifted GMR metric \eq{GMRsln} resembles a generalization of the Garfinkle-Vachaspati construction, using not one but two vectors $k^{\mu}$, $l^{\mu}$. We take 
\bea
k^\mu \d_\mu &=& \frac{\d}{\d u} \,, \qquad l^\mu \d_\mu ~=~ \frac{\d}{\d z}
\eea
so both are Killing, but only $k$ is null. Lowering the indices with the metric \eq{GMRsln}, we obtain the one-forms
\bea
k_\m dx^{\m} &=& -\frac{1}{2H} ( dv + \beta )  \,, \qquad l_\m dx^{\m} ~=~ dz \,.
\eea
Then one can see that, absorbing a factor of 2, the deformation in Section \ref{SectSSgrn} takes the form
\bea
g_{\m\n} &=& {\bar g}_{\m\n} + \Phi \, k_\mu l_\nu \, ,\qquad
C^{(2)}={\tilde C}^{(2)}+\frac{1}{2}\Phi~ 
k_\m dx^{\m}\wedge l_\n dx^{\n}
\eea
where $\Phi$ is harmonic in the background and independent of $u$ and $z$.

The presence of the two-form potential $C^{(2)}$, and the use of two Killing vectors, differentiate the current setup from the one studied by Garfinkle-Vachaspati. Nevertheless it seems tempting to wonder whether a similar mechanism underlies our work, and whether this observation may be useful in other scenarios. We leave a full investigation of this for future work.

\section{Discussion} \label{sec:disc} 

In this paper we have presented a deformation which adds a travelling wave to the GMR construction~\cite{\GMR}, trivially lifted to 10D. 
Our construction generalizes the linearized solutions of \cite{Mathur:2011gz,Mathur:2012tj}, both to the nonlinear level and to a large class of different backgrounds. 

In these earlier works the deformation of the present paper was identified as corresponding to the action of a U(1) generator $J^\a_{-n}$ of the D1-D5 orbifold CFT, by connecting to the work of Brown and Henneaux~\cite{Brown:1986nw}. 
The background solution used in \cite{Mathur:2011gz,Mathur:2012tj} has, in a certain regime of parameters, a large AdS region.
The deformation arises at the boundary of this region and
appears to be related to the `singleton' (or `doubleton') representations that lie at the boundary of AdS \cite{Dirac:1963ta,*Flato:1978qz,*Gunaydin:1986fe}.

Applied to a general GMR background, our deformation is given in terms of four functions $\Phi_\a$ which satisfy the wave equation on the background. We have studied a large class of two-charge D1-D5 backgrounds~\cite{\lmAdS}, and the particular examples of a family of D1-D5-P backgrounds~\cite{Lunin:2004uu,Giusto:2004id,*Giusto:2004ip}. In the latter case we have found the functions $\Phi_\a$ explicitly. In each case, the deformed solution preserves the regularity of the background.

The results presented in the current paper offer many opportunities for future research. Most obviously, it is natural to ask whether our deformation can be applied to more general background solutions, as conjectured in Section \ref{SectBW}. We have also noted connections to pp-wave solutions and the Garfinkle-Vachaspati construction in Section \ref{sec:pp-wave-gv}. 
From the D1-D5 CFT point of view, it is also natural to look for the nonlinear deformations corresponding to the other bosonic generators of the symmetry algebra, namely the Virasoro generators $L_n$ and the SU(2) R-symmetry currents ${\cal J}^a_n$. 

\section*{Acknowledgements}

We thank Borun Chowdhury and Stefano Giusto for discussions.
The work of SDM and DT  was supported in part by DOE grant DE-FG02-91ER-40690.

\begin{appendix}

\section{Review of the GMR construction}
\label{AppGMR}

In this appendix we review the GMR notation and equations. 
In~\cite{Gutowski:2003rg} it was shown that locally, any supersymmetric solution of minimal 6D supergravity may be written in the form which we now review. The metric takes the form\footnote{Our conventions are related to those of \cite{Gutowski:2003rg} via $u=\sqrt{2}\,v_{GMR}$, $v=\sqrt{2}\,u_{GMR}$, $\beta=\sqrt{2}\,\beta_{GMR}$, $\omega=\sqrt{2}\,\omega_{GMR}$, and $\hat{u}=+$, $\hat{v}=-$.}:
\bea\label{GMRsln-1}
ds^2&=&-e^{\hat v}e^{\hat u}+Hds_4^2 
 \,, \qquad e^{\hat v}~=~H^{-1}(dv+\beta) \,,\quad
e^{\hat u}=du+\omega+\frac{\cF H}{2}e^{\hat v} \,
\eea
Here $ds_4^2$ is a metric on a 4D base manifold, $H$ and $\cF$ are functions and $\beta$ and $\omega$ are 1-forms on the base. All of these can be $v$--dependent and must satisfy further conditions reviewed below.

The base manifold must admit an almost hyper-K\"ahler structure with almost complex
structures $J^i$ ($i=1,2,3$) which are anti-self-dual 2-forms on the base. These must obey
\be
\label{eq:cdJ}
{\tilde{d}} J^i = \partial_v \big(\beta \wedge J^i \big) \,
\ee
where $\tilde d$ is the exterior derivative on the base, see \eq{eq:dtilde}.
Recalling the notation
\bea
D\Phi ~=~ \tilde{d}\Phi-\beta\,\d_v\Phi \,,
\eea
and defining
\be
\label{eqn:psidef}
\psi = {H \over 16} \epsilon^{ijk} (J^i)^{pq} ({\dot{J^j}})_{pq} J^k \,,
\ee
the 3-form field strength may be written as $(~\dot{} = \d/\d_v)$
\bea
\label{eq:Gexpand}
F &=& {1 \over 2} \star_4 \left(D H + H  {\dot{\beta}} \right)
+e^{\hat u} \wedge \left(H \psi - {1 \over 2} (D \omega)^{\hat v}\right)
\nn
&+& {1 \over 2} H^{-1} e^{\hat v} \wedge D \beta - {1 \over 2} e^{\hat u} \wedge e^{\hat v}
\wedge \left(H^{-1} D H+ {\dot{\beta}} \right).
\eea
Then the $C^{(2)}$ can be written as\footnote{
To see this, one should first observe that the term $\frac{1}{2} e^{\hat v}\wedge e^{\hat u}$ in $C^{(2)}$ accounts 
for $F_{u\mu\nu}$ in (\ref{eq:Gexpand}). The remaining part of $C^{(2)}$ can only contain differentials $dv$ and $dx^i$, and (\ref{GexpA}) gives the most general form of such contributions.}
\bea\label{GexpA}
C^{(2)}&=&\frac{1}{2} e^{\hat v}\wedge e^{\hat u} +
e^{\hat v}\wedge {\cal A}+\sigma_2 \,.
\eea
We now summarize the equations coming from the Bianchi identity and Einstein equations.
In terms of the self-dual 2-form
\be
 {\cal G}^+ \equiv H^{-1} \left( (D \omega)^+ + \frac{1}{2} {\cal F}
 D \beta \right) \,,
\ee
the Bianchi identity $dG=0$ gives the following constraints:
\be
 \label{eqn:biana} D \left(\star_4 (D H + H
 {\dot{\beta}}) \right) + D \beta \wedge {\cal G}^+ =0 
\ee
and
\be
\label{eq:bianb}
 \tilde{d} \left( {\cal G}^+ + 2 \psi \right) = \partial_v \left[ \beta
 \wedge \left( {\cal G}^+ + 2\psi \right) + \star_4 \left( D H + H \dot{\beta} \right)
 \right] .
\ee
Finally, the Einstein equation has one nontrivial component, the ${\hat v}{\hat v}$ component. Defining
\be
L =   {\dot{\omega}} + \frac{1}{2} \cF {\dot{\beta}} -{1 \over 2} D \cF \,,
\ee
this component reduces to
\bea
\label{eq:ee}
 \star_4 D (\star_4 L) &=&
 \frac{1}{2} H h^{mn} \partial_v^2 (H h_{mn})
+{1 \over 4}  \partial_v (Hh^{mn}) \partial_v (Hh_{mn})
- 2{\dot{\beta}}_m L^m
\nn
&+& \frac{1}{2} H^{-2}\left((D \omega)^- - 2 H \psi \right)^2
-{1 \over 2} H^{-2} \left( D \omega + \frac{1}{2} \cF \beta \right)^2.
\eea

\section{Equations of motion for the deformation}
\label{AppDeform}

This appendix is dedicated to the derivation of the solution (\ref{GMRdfm}), (\ref{GMRLapl}). We start with the ansatz (\ref{GMRdfm}) and verify the equations of motion and equations for the Killing spinors. We show that the deformation (\ref{GMRdfm}) gives a supersymmetric solution of type IIB supergravity, as long as the original geometry (\ref{GMRsln}) solves the GMR equations and each of the four scalars $\Phi_\alpha$ obeys equation (\ref{GMRLapl}). To simplify the discussion, we turn on only one deformation $\Phi=\Phi_1$, but the extension to several nontrivial $\Phi_\alpha$ is straightforward. 

\subsection{Equations for the field strength}

We begin with imposing the ansatz (\ref{GMRdfm}) and deriving the equations for the field strength. 
This will lead to equation (\ref{GMRLapl}) for the scalar $\Phi$.

In minimal supergravity, the two--form $C^{(2)}$ satisfies Maxwell's equations:
\bea \label{eq:b1}
d(*F)~=~0 \,, \qquad F ~=~ dC^{(2)} \,.
\eea
To check this equation, we begin with evaluation of $F$ :
\bea\label{dBFrames}
F&=&\frac{1}{2}d\Phi\wedge dz \wedge e^{\hat v} +
\left\{\frac{1}{2} \left[de_{(0)}^{\hat u} \wedge e^{\hat v}-
e^{\hat u} \wedge de^{\hat v}\right]+
d(e^{\hat v}\wedge {\cal A}+\sigma_2)\right\}
\eea
Here we defined
\bea
e_{(0)}^{\hat u}=du+\omega+\frac{\cF}{2}e^{\hat v}
\eea
The terms in the curly brackets in (\ref{dBFrames}) do not contain derivatives of $\Phi$, and $\Phi$  dependence appears only through $\Phi dz$ in $e^{\hat u}$. This guarantees that the Hodge dual of this bracket depends on $\Phi$ only through $\Phi dz$ in $e^{\hat u}$ as well, and since the dual of the terms in the curly brackets contains an overall factor of $dz$, the following expression does not contain $\Phi$:
\bea
*_{10}\left\{\frac{1}{2} \left[de^{\hat u}_{(0)} \wedge e^{\hat v}-
e^{\hat u} \wedge de^{\hat v}\right]+
d(e^{\hat u} \wedge {\cal A}+\sigma_2)\right\} \,.
\eea
Therefore, given that the undeformed $C^{(2)}$ satisfies its equation, the equation \eq{eq:b1} reduces to 
\bea \label{eq:c2eqn}
d(*_{10}\left[d\Phi\wedge dz \wedge e^{\hat v}\right]) ~=~ 0.
\eea
Using the relation\footnote{We recall the notation introduced in the Appendix \ref{AppGMR}: 
${\tilde d}f=\d_i fdx^i$, $Df={\tilde d}f-\beta \d_v f$.
}
\bea \label{eq:d-to-D}
d\Phi\wedge e^{\hat v}=({\tilde d}-\beta \d_v)\Phi \wedge e^{\hat v}+
\d_v\Phi (dv+\beta)\wedge e^{\hat v}=({\tilde d}-\beta \d_v)\Phi \wedge e^{\hat v}\equiv 
D\Phi \wedge e^{\hat v},
\eea
equation \eq{eq:c2eqn} can be rewritten as
\bea\label{Lapl4DAp}
d(H[*_{4} D\Phi]\wedge e^{\hat v})=0,
\eea
where Hodge dual is taken with respect to the four--dimensional geometry of the base. This gives equation (\ref{GMRLapl4}), and now we will prove that (\ref{Lapl4DAp}) can be viewed as the wave equation for $\Phi$ on six--dimensional space formed by $u,v,x^i$. Indeed, application of the D'Alembert operator to a $u$--independent field gives
\bea
d*_6\left[D\Phi+\d_v\Phi He^{\hat v}\right]&=&d[\frac{1}{2}(*_4D\Phi)\wedge
e^{\hat v}\wedge e^{\hat u}+\d_v\Phi He^{\hat v}\wedge \mbox{Vol}_4]\nonumber\\
&=&\frac{1}{2}d[(*_4D\Phi)e^{\hat v}]du
\eea
We used the fact that the six--form on 6D space must contain $du$, which can only come from $e^{\hat u}$. This demonstrates the equivalence of equations (\ref{GMRLapl}) and (\ref{GMRLapl4}). 

To summarize, we showed that equation (\ref{Lapl4DAp}) guarantees that the RR form in the ansatz (\ref{GMRdfm}) satisfies the correct equation of motion.

\subsection{Killing spinor equations}

We next check that the geometry (\ref{GMRdfm}) admits a Killing spinor.
Since the deformation is in 10D, we analyze the Killing spinor equations in type IIB supergravity. 
Starting with the general form of these equations \cite{\Grana}, and setting all matter fields except $C^{(2)}$ to zero, 
we find the equations 
\bea\label{RRKilling}
&&{\not F}\eps^*~=~0 \,,\qquad\qquad\qquad\qquad\qquad  {\not F}~\equiv~ F_{abc}\gamma^{abc} \,,\nonumber\\
&&\nabla_a\eps+\frac{i}{96}(-\gamma_a {\not F}-
2{\not F}\gamma_a)\eps^*~=~0 \,.
\eea
In this subsection the letters from the beginning of the alphabet refer to the orthonormal frame, as do all hatted indices. For example,
\bea
\{\gamma_a,\gamma_b\}=2\eta_{ab} \,.
\eea
Since ${\not F}$ is a real matrix, it is convenient to combine equations (\ref{RRKilling}) with their complex conjugates and rewrite the resulting relations in terms of a new complex spinor
\bea
\eta=\eps+i\eps^*.
\eea
This leads to the equations with real coefficients:
\bea\label{dltno}
&&{\not F}\eta=0 \,,\\
\label{grvno}
&&\nabla_a\eta+
\frac{1}{96}(-\gamma_a {\not F}-
2{\not F}\gamma_a)\eta=0 \,.
\eea
We assume that equations (\ref{dltno}), (\ref{grvno}) are satisfied by the background with $\Phi=0$ and demonstrate 
that the deformed configuration also satisfies these relations. 

To verify equation (\ref{dltno}) from the dilatino variation, 
we recall that the undeformed background satisfied the 
projection:
\bea\label{PlusProj}
\gamma^{\hat v}\eta&=&0 \,.
\eea
To verify (\ref{dltno}) in the deformed case, we rewrite (\ref{dBFrames}) as 
\bea \label{eq:deformed-F}
F=F^{(0)}+
\frac{1}{2}d\Phi\wedge e^{\hat z}\wedge e^{\hat v}
\equiv F^{(0)}+F^{(1)},
\eea
and notice that the frame components of $F^{(0)}$ do not contain $\Phi$. The projection (\ref{PlusProj})
ensures that $F^{(1)}$ disappears from equation (\ref{dltno}),
so the equation for the dilatino is not affected by the deformation.  

Next we check equation (\ref{grvno}) for the gravitino. The projection (\ref{PlusProj}) implies that ${\not F}^{(1)}\eps=0$, so (\ref{grvno}) can be rewritten as
\bea
\label{grvno1}
&&\nabla_a\eta-\frac{1}{96}(\gamma_a {\not F}^{(0)}+
2{\not F}^{(0)}\gamma_a)\eta
-\frac{1}{48}{\not F}^{(1)}\gamma_a\eta=0
\eea
To verify this equation, we need the spin connection $\omega_{a,bc}$. Our conventions are
\bea
de^{a}+ \T{\omega}{a}{b} \wedge e^{b} &=& 0 \,, \qquad \omega_{ab} ~=~ - \omega_{ba} \,, \qquad \omega_{ab} 
~=~ \omega_{c,ab} \, e^c \,,
\eea
\bea
de^{a} &=& \ha \Gamma^a{}_{,bc} \, e^b \wedge e^c \,, \qquad \omega_{c,ab} ~=~ \ha 
\left[ \Gamma_{c,ab} + \Gamma_{b,ac} - \Gamma_{a,bc} \right] \,,
\eea

and the covariant derivative is
\bea
\nabla_a\eta ~=~ \left(e_a^\mu \d_\mu +\frac{1}{4}\omega_{a,bc}\gamma^{bc}\right)\eta \,.
\eea
The deformation introduces the following new components of the spin connection:
\bea \label{eq:spin-con}
\omega_{{\hat v},{\hat i}{\hat z}} &=& \omega_{{\hat i},{\hat v}{\hat z}} ~=~ - \omega_{{\hat z},{\hat v}{\hat i}}
~=~ \frac{1}{4} (D\Phi)_{\hat i} \,, 
\qquad\quad  \omega_{{\hat v},{\hat v}{\hat z}} ~=~ \ha H \, \d_v \Phi \,.
\eea
The projection (\ref{PlusProj}) implies that only the $a={\hat v}$ component of equation \eq{grvno1} is affected by the deformation. The new terms in this component are
\bea
\frac{1}{2}\omega_{{\hat v},{\hat i}{\hat z}}\gamma^{{\hat i}{\hat z}} \eta -\frac{1}{48}{\not F}^{(1)}\gamma_{\hat v}\eta 
&=& \frac{1}{8} (D\Phi)_{\hat i}\gamma^{{\hat i}{\hat z}} \eta 
- \frac{1}{16} (D\Phi)_{\hat i}\gamma^{{\hat i}{\hat z}{\hat v}}\gamma_{\hat v}\eta  ~=~ 0 \,,
\eea
where in the first step we used \eq{eq:deformed-F} with \eq{eq:d-to-D}, and in the second step we moved $\gamma^{\hat v}$ through $\gamma_{\hat v}$ and using the projection (\ref{PlusProj}).

\subsection{Einstein equation}

Having solved the Killing spinor equations and $C^{(2)}$ field equation, it remains to check the the ${\hat v}{\hat v}$ component of the Einstein equation~\cite{Gauntlett:2002fz}.

Using \eq{eq:spin-con} with the standard formulae for the curvature 2-form $\T{\mathcal{R}}{a}{b}$
\bea
\T{\mathcal{R}}{a}{b} &=& \T{d\omega}{a}{b} + \T{\omega}{c}{\,c} \wedge \T{\omega}{c}{\,b} 
\eea
and the Riemann tensor in the orthonormal frame,
\bea
\ha \T{R}{a}{bcd} e^{c} \wedge e^{d} &=& \T{\mathcal{R}}{a}{b} \,,
\eea
we find
\bea
\qquad\quad  R_{\hat v \hat v} &=& \frac{1}{4} F_{\hat v ab} F_{\hat v}{}^{ab}  \,~=~\, \frac{1}{8} (D\Phi)_{\hat i} (D\Phi)_{\hat i} \,. 
\eea
Thus the deformed configuration satisfies the IIB Killing spinor equations and Einstein equation.

\section{Laplace equation for D1--D5 geometries}
\label{app:laplace}

\subsection{Uniqueness of the solution}
\label{AppUnq}

While discussing the deformation of supersymmetric geometries in Section \ref{SectSSgrn}, we encountered the generalized Laplace equation (\ref{D1D5Lapl}). To describe regular geometries, we are looking for solutions satisfying conditions (a)--(c) in page \pageref{PageCndPhi}, and in this section we will demonstrate that if a solution satisfying these requirements exists, then it is unique. Our proof will be very similar to the one used for the standard Laplace equation. To avoid unnecessary indices, we will focus 
on one of the fields $\Phi\equiv \Phi_1$. 

Since the four--dimensional base is flat and $v$--independent, 
one can rewrite equation (\ref{D1D5Lapl}) in a more explicit form:
\bea\label{ScalLapl}
(\d_i-B_i\d_v)\left[
(\d_i-B_i\d_v)\right]\Phi=0
\eea
Let us start with a periodic function $f(v)$ and assume that (\ref{ScalLapl}) admits two solutions, which are regular in the interior and approach $f(v)$ at infinity. Then the difference of these two solutions goes to zero at infinity and remains regular in the interior, and we will demonstrate that the only solution with such properties is $\Phi=0$. 

Let us assume that there exists a solution $\Phi$ of (\ref{ScalLapl}), which approaches zero at infinity. Since variable $v$ is periodic and equation (\ref{ScalLapl}) is invariant under translations in $v$, different Fourier harmonics decouple, and without loss of generality we assume that $\Phi=g(x)e^{-ipv}$. As demonstrated in \cite{\lmAdS,\lmm}, vector field $B$ remains regular everywhere away from the one--dimensional profile $x_i=F_i(w)$ (see 
equation (\ref{StrProfile})). Let us draw a sphere at some large value of $r=R$ and surround the profile by a tube with radius $\eps$. In the region between the tube and the sphere, function $\Phi$ and field $B_i$ remain regular, so we can perform standard manipulations with integration 
by parts. Multiplying (\ref{ScalLapl}) by ${\overline\Phi}$ and integrating over the region between the tube and the sphere, we find
\bea
0&=&\int_D d^4 x {\overline\Phi} \left[(\d_i-B_i\d_v)(\d_i-B_i\d_v)\right]\Phi\nonumber\\
&=&
-\int_D d^4 x |(\d_i-B_i\d_v)\Phi|^2
+\int_{\d D} dS_i {\overline\Phi} (\d_i-B_i\d_v)\Phi\\
&&-\int_D d^4 x {\overline\Phi} B_i(\d_i-B_i\d_v)\d_v\Phi
-\int_D d^4 x \d_v{\overline\Phi} B_i(\d_i-B_i\d_v)\Phi\nonumber \,.
\eea
In particular, we focus in the real part of the last equation:
\bea\label{Jun18}
0&=&-\int_D d^4 x |(\d_i+ipB_i)g|^2
+\int_{\d D} dS_i {\bar g} \d_ig \,.
\eea
The second term has a well defined limit as 
$\eps$ goes to zero and $R$ goes to infinity, and it goes to zero in this limit\footnote{This follows from regularity of $\Phi$ and boundary condition $f(\infty)=0$.}. Then equation (\ref{Jun18}) implies that 
\bea\label{CovDerZer}
(\d_i+ipB_i)g=0 \,.
\eea
This is the set of four equations, which must satisfy integrability conditions
\bea
(\d_j+ipB_j)(\d_i+ipB_i)g=0 \,.
\eea 
Antisymmetric part of this tensor gives a relation which does not involve derivatives of $f$,
\bea
p(\d_i B_j-\d_j B_i)g=0,
\eea
and since $B_idx^i$ is not a pure gauge, we conclude that either $g=0$ or $p=0$. In the latter case, equation (\ref{CovDerZer})
implies that $g$ is a constant, and since it must vanish at infinity, we again conclude that $g=0$. 

We proved that any function $\Phi=g(x)e^{-ipv}$, which satisfies equation (\ref{ScalLapl}), approaches zero at infinity and remains finite everywhere must be equal to zero. As we already discussed, this proves the uniqueness theorem for equation (\ref{ScalLapl}) with requirements (a)--(c) listed in page \pageref{PageCndPhi}. 


\subsection{Explicit solutions for the special case}
\label{AppGreen}

In this subsection we will focus on deformation (\ref{MMConj}) constructed in Section \ref{SectMM}. In particular, to construct a dual of a given state in the CFT, we need to recover function $\Phi$ from the boundary data $f(v)$. Although this can be done by expanding $f(v)$ in the Fourier series (\ref{fOfV}) and writing the corresponding (\ref{PhiFourier}), it is desirable to have a more explicit relation. In general, the solutions of linear PDEs can be recovered from the boundary data by using the Green's function, and in this subsection we will construct the Green's function for our problem which  would lead to (\ref{MainGreen}). 

Consider the boundary value problem for $\Phi(r,v)$:
\bea\label{Nabla6}
\nabla_6^2 \Phi=0,\qquad \Phi(\infty,v)=f(v).
\eea
Here $\nabla_6^2$ is the D'Alembert operator on the six--dimensional space with metric 
\bea
ds_6^2=-\frac{1}{H}\left[du+A\right]\left[dv+B\right]+
Hf\left[\frac{dr^2}{r^2+a^2}+d\theta^2\right]
+H\left[r^2c_\theta^2 d\psi^2+(r^2+a^2)s_\theta^2d\phi^2\right] \,. \nonumber 
\eea
The last subsection guarantees that equations (\ref{fOfV}), (\ref{PhiFourier}) give the 
unique solution with given asymptotics, and now we will rewrite the relation between $f(v)$ and $\Phi$ in a more explicit form. To find such relation, we first look at
\bea
f(v)=\sum_{n=-\infty}^\infty c_n e^{-inv/R_y},\qquad
c_n=\frac{1}{2\pi R_y}\int_0^{2\pi R_y} dv' e^{inv'/R_y}f(v')\,.
\eea 
The corresponding solution in the bulk is
\bea\label{Jun30a}
\Phi&=&\sum_{n=0}^\infty c_n q^n e^{-inv/R_y}+\sum_{n=0}^\infty c_{-n} q^n e^{inv/R_y}-c_0\equiv \Phi_++\Phi_--c_0\,.
\eea
Here we defined the expansions $\Phi_+$, $\Phi_-$ and introduced a convenient variable
\bea
\label{DefQser}
q&\equiv&\left(\frac{r^2}{r^2+a^2}\right)^{1/2}<1.
\eea
Function $\Phi_+$ can be evaluated using the geometric series:
\bea
\Phi_+&=&\sum_{n=0}^\infty c_{n} q^n e^{-inv/R_y}=
\int_0^{2\pi R_y}\frac{dv'}{2\pi R_y}f(v')\left[1-q 
e^{-i(v-v')/R_y}\right]^{-1} \,. \nonumber
\eea
Combining this with similar expression for $\Phi_-$ and with definition of $c_0$, we find 
the  relation between $\Phi$ and $f(v)$:
\bea\label{GrnRslt}
\Phi=\int_0^{2\pi R_y}\frac{dv'}{2\pi R_y}f(v')G(r;v,v'),\quad
G(r;v,v')=\frac{1-q^2}{
1+q^2-2q\cos\frac{v-v'}{R_y}} \,.
\eea 
The algebra leading to (\ref{GrnRslt}) is straightforward:
\bea
G(r;v,v')&=&\frac{1}{1-q e^{-i(v-v')/R_y}}+
\frac{1}{1-q e^{i(v-v')/R_y}}-1\nonumber\\
&=&\frac{2-2q\cos\frac{v-v'}{R_y}}{
1-2q\cos\frac{v-v'}{R_y}+q^2}-1=
\frac{1-q^2}{
1+q^2-2q\cos\frac{v-v'}{R_y}} \,. \nonumber
\eea

As a consistency check, we demonstrate that equation (\ref{GrnRslt}) recovers the boundary value. As $r$ goes to infinity, $q$ approaches one, and $G(r;v,v')$ approaches a periodic array of delta functions: it vanishes unless $v=v'+2\pi nR_y$, and the integrals involving periodic functions $e^{inv'/R_y}$ are correct\footnote{We evaluate the integral only for $n\ge 0$ by introducing a complex variable $z=e^{iv'/R_y}$ and using the residue theorem. The integral for $n<0$ can be evaluates in the same way, but one has to use $z=e^{-iv'/R_y}$.}:
\bea
I_n&=&
\int_0^{2\pi R_y}\frac{dv'}{2\pi R_y}\frac{e^{inv'/R_y}(1-q^2)}{
1+q^2-2q\cos\frac{v-v'}{R_y}}=
\int_0^{2\pi R_y}\frac{dv'}{2\pi R_y}\frac{e^{in(v'+v)/R_y}(1-q^2)}{
1+q^2-2q\cos\frac{v'}{R_y}}\nonumber\\
&=&e^{inv/R_y}\int_0^{2\pi R_y}\frac{dv'}{2\pi R_y}\frac{e^{inv'/R_y}(1-q^2)}{
1+q^2-2q\cos\frac{v'}{R_y}}=
e^{inv/R_y}\oint\frac{dz}{2\pi i}\frac{z^{n-1}(1-q^2)}{
1+q^2-q(z+z^{-1})}\nonumber\\
&=&e^{inv/R_y}\mbox{res}\left[\frac{z^{n-1}(1-q^2)}{
1+q^2-q(z+z^{-1})},z=q\right]=
e^{inv/R_y}\frac{q^{n-1}(1-q^2)}{-q+q^{-1}}\rightarrow 
e^{inv/R_y}\nonumber
\eea

To summarize, the unique solution $\Phi(r,v)$ corresponding to an arbitrary boundary value $f(v)$ is given by equation (\ref{GrnRslt}).

\section{Charges of the deformed D1--D5--P solution} \label{app:Charges}

To extract the charges of the solution \eq{PhiAlpha-SF}--\eq{DfrSpFl-C2}, we 
need to extend the diffeomorphism (\ref{FOfVFlow-2}) to finite values of the radial coordinate\footnote{The subleading terms omitted in (\ref{DiffToInf}) do not contribute to the charges.},
\bea\label{DiffToInf}
z'_\alpha=z_\alpha-\frac{1}{2}\int \Phi_\alpha dv,\quad
u'=\la\left[u+\frac{1}{4}\int \Phi_\alpha \Phi_\alpha dv\right]+\dots,\quad
v'=\frac{v}{\la}+\dots,
\eea
and to rewrite the metric (\ref{DfrSpFl}) in terms of coordinates $z'_\alpha$ $u'$, $v'$. 
Going to large values of $r$, while keeping only the flat contribution and the leading nontrivial terms in  $g_{u'\mu}$
and $g_{v'\mu}$, one finds
\bea
ds^2&=&-\left(1-\frac{Q}{r^2}\right)du'dv'
+dr^2+r^2(d\theta^2+c_\theta^2d\psi^2+s_\theta^2 d\phi^2)
+dz'_\alpha dz'_\alpha\nonumber\\
&&+\frac{\la Q}{r^2}h_\alpha dv' dz'_\alpha+\frac{Q\la^2 }{4r^2}h_\alpha h_\alpha (dv')^2
+\frac{\la^2 Q_p}{r^2}(dv')^2\\
&&-\frac{2Qa\la}{r^2}[-mc_\theta^2 d\psi+
(m+1)s_\theta^2 d\phi]
dv'
-\frac{Qa\la\eta}{r^2}[c_\theta^2 d\psi+s_\theta^2 d\phi]
\left[\frac{du'}{\la^2}-\frac{h_\alpha h_\alpha}{4} dv'- dv'\right]+\dots\nonumber
\eea
To read off the charges of the solution, we introduce coordinates 
$t'=\frac{u'+v'}{2}$, $y'=\frac{u'-v'}{2}$ and look at the leading corrections of $g_{t'\mu}$:
\bea
g_{t't'}&=&-\left[1-\frac{Q}{r^2}-\frac{\la^2Q_p}{r^2}-\frac{\la^2Q}{4r^2}h_\alpha h_\alpha\right],\quad
g_{t'\alpha}=\frac{\la Q}{2r^2}h_\alpha,\quad 
g_{t'y}=-\frac{\la^2Q_p}{r^2}-\frac{\la^2Q}{4r^2}h_\alpha h_\alpha\nonumber\\
\label{ChGtPhi}
g_{t'\phi}&=&-\frac{\la Qa}{r^2}s_\theta^2\left[m+1-\frac{\eta}{2}\left\{1+\frac{1}{4}h_\alpha h_\alpha-\frac{1}{\la^2}\right\}\right],
\\
g_{t'\psi}&=&\frac{\la Qa}{r^2}c_\theta^2\left[m-\frac{\eta}{2}\left\{1+\frac{1}{4}h_\alpha h_\alpha-\frac{1}{\la^2}\right\}\right] \,.
\nonumber
\eea
The momenta and mass of the solution are given by\footnote{We are using the fact that $c_0^\alpha=0$.}
\bea\label{PyChrg}
P_\alpha&=&-\frac{\pi}{4G_N}\int dy' r^2\delta g_{t'\alpha}=
-\frac{\pi\la}{4G_N}\int dy' \frac{1}{2}f_\alpha=0,\nonumber\\
P_{y'}&=&-\frac{\pi}{4G_N}\int dy' r^2\delta g_{t'y'}=
\frac{\pi\la^2}{4G_N}\left[2\pi R Q_p+
\frac{Q}{4}\int_0^{2\pi R} dy' h_\alpha h_\alpha\right]
\\
M&=&-\frac{\pi}{4G_N}\int dy' r^2\delta g_{t't'}=
\frac{\pi\la^2}{4G_N}\left[2\pi R Q +2\pi R Q_p+\frac{Q}{4}\int_0^{2\pi R} dy' h_\alpha h_\alpha\right]
\nonumber\\
&=&\frac{\pi^2 R Q\la^2}{2G_N}+P_{y'} \,. \nonumber
\eea
Here $G_N$ is the six--dimensional Newton's constant. 

As expected for a BPS state, introduction of momentum along $y$ direction shifts mass by $P_{y'}$.   Recalling the expressions for the undeformed $P^{(0)}_y$ and $Q_p$ in terms of integers $n_p,n_1,n_5,m$,
\bea\label{Jul23}
P^{(0)}_y=\frac{n_p}{R},\quad Q_p=-\gamma_1\gamma_2=a^2 m(m+1)\equiv a^2 \frac{n_p}{n_1n_5},
\eea
we can rewrite (\ref{PyChrg}) as
\bea\label{DfrmPy}
P_{y'}&=&
\frac{n_1n_5}{R}\left[m(m+1)+\frac{Q}{4a^2}\frac{1}{2\pi R}\int_0^{2\pi R} dy' h_\alpha h_\alpha\right] \,.
\eea
For future reference, we notice that comparison of  (\ref{PyChrg}) with (\ref{Jul23}) leads to the relation
\bea\label{QSqrn15}
\frac{\pi^2\la^2 Ra^2}{2G_N}=\frac{n_1n_5}{R}\quad
\Rightarrow\quad
\frac{\pi^2Q^2}{2G_N}=n_1n_5, 
\eea
which can be verified by evaluating the RR charges of the solution \eq{PhiAlpha-SF}--\eq{DfrSpFl-C2}.

Finally, we use (\ref{ChGtPhi}) and (\ref{QSqrn15}) to extract the angular momenta:
\bea\label{DfrmAngMom}
J_\phi&=&-\frac{\pi}{8G_N}\int_0^{2\pi R} 
dy' r^2\frac{\delta g_{t'\phi}}{\sin^2\theta}=
\frac{\pi Qa\la}{8G_N}\int_0^{2\pi R}dy'
\left[m+1-\frac{\eta}{2}\left\{1+\frac{1}{4}h_\alpha h_\alpha-\frac{1}{\la^2}\right\}\right]
\nonumber\\
&=&
\frac{n_1n_5}{2}\left[m+1\right],\\
J_\psi&=&
-\frac{n_1n_5}{2}m \,.
\eea
Here we used expression (\ref{LmbdBoost}) for $\la^{-2}$. Let us summarize the results:
\bea\label{sumChrgAp}
&&P_\alpha=0,\qquad P_{y'}=
\frac{n_1n_5}{R}\left[m(m+1)+\frac{Q}{8\pi R a^2}
\int_0^{2\pi R} dy' {h_\alpha h_\alpha}\right],
\nonumber\\
&&J_\phi=\frac{n_1n_5}{2}\left[m+1\right],\quad
J_\psi=
-\frac{n_1n_5}{2}m.
\eea
As discussed in Section \ref{sec:MM_cft_map}, these charges agree with CFT expectations.

To map the geometries into states in the dual CFT, one should evaluate the charges in the AdS region rather than in asymptotically flat space. Such calculation is possible only if the geometry contains a long AdS throat, where $a^2\ll r^2\ll Q$. In particular, this implies that $\eps\equiv\frac{a^2}{Q}\ll 1$. Let us briefly discuss the extraction of charges from the AdS region and demonstrate that expressions (\ref{sumChrg}) are recovered. 

To compare with CFT picture, one should take the limit $\eps\rightarrow 0$, while keeping the radius of AdS space, $\sqrt{Q}$, fixed. Then relation (\ref{MMSpFlaRy}) implies that 
$R_y$ goes to infinity, so it is convenient to define a new coordinate ${\tilde y}=y/R_y$, whose period remains unchanged. To have the standard form of the AdS asymptotics in the small $\eps$ limit, we fix ${\tilde t}=t/R_y$ as well.
To summarize, we are taking the limit
\bea\label{AdSLimit}
\eps=\frac{a^2}{Q}=\frac{Q}{R_y^2}\rightarrow 0:\quad
\mbox{fixed}\ Q,\ {\tilde u}=\frac{u}{R_y},\ 
{\tilde v}=\frac{v}{R_y},\ z^\alpha,\ 
{\tilde \Phi}_\alpha=\frac{\sqrt{Q}}{a}\Phi_\alpha,\
{\tilde f}_\alpha=\frac{\sqrt{Q}}{a}f_\alpha \,.
\eea
The scaling of $\Phi_\alpha$ is introduced to get finite contribution to $P_{y'}$ in (\ref{DfrmPy}) as $\eps$ goes to zero. 
Equation (\ref{LmbdBoost}) implies that $\la\rightarrow 1$ in the limit, then $y=y'$, $t=t'$, $R_y=R$. Notice that there is still a nontrivial shift of the $z_\alpha$ coordinates in (\ref{FOfVFlow-2}), and it leads to the momentum charge. The angular momenta are not affected by the limit, and AdS momentum becomes
\bea
P_{\tilde y}=
{n_1n_5}\left[m(m+1)+\frac{1}{8\pi}\int_0^{2\pi} d{\tilde y} {{\tilde f}_\alpha {\tilde f}_\alpha}\right],
\eea

Although we evaluated the charges by going to the flat asymptotics, the same expressions can be extracted from the AdS region. In the AdS throat, $a\ll r\ll \sqrt{Q}$, the leading contribution to the metric (\ref{DfrSpFl}) becomes
\bea
ds^2&=&-Q\left[\left(\frac{rR_y}{Q}\right)^2
\left\{d{\tilde u}+
{\tilde f}_\alpha({\tilde v}) \frac{dz^\alpha}{R^2_y}\right\}d{\tilde v}+
\frac{d{r}^2}{{r}^2}\right]+Qd\Omega_3^2+
dz^\alpha dz^\alpha,
\eea
and to get the standard AdS asymptotics,  
${\tilde r}=\frac{rR_y}{Q}$ will be kept fixed in the limit (\ref{AdSLimit}). Notice that the terms containing 
${\tilde f}_\alpha({\tilde v})$ disappear in the limit (\ref{AdSLimit}), and, as expected,  the metric reduces to  $AdS_3\times S^3\times T^4$. 

To read off charges in the AdS region, one should take the limit 
(\ref{AdSLimit}) in (\ref{DfrSpFl}), including the subleading terms in $1/{\tilde r}^2$, and rewrite the result in the primed coordinates (\ref{DiffToInf}). We begin with rewriting (\ref{DiffToInf}) in terms of rescaled variables defined in (\ref{AdSLimit}):
\bea
{z}_\alpha=
{z}'_\alpha+
\frac{1}{2}\int {\tilde\Phi}_\alpha d{\tilde v},\quad
{\tilde u}={\tilde u}',\quad {\tilde v}={\tilde v}' \,.
\eea
Introducing the rescaled variables in in (\ref{DfrSpFl}) and 
sending $\eps$ to zero, one finds
\bea\label{AdSDiff}
ds^2&=&Q\left[-{\tilde r}^2d{\tilde u}d{\tilde v}-
\frac{1}{4}(d{\tilde u}+d{\tilde v})^2+
\frac{d{\tilde r}^2}{{\tilde r}^2+1}\right]+
\left(dz'_\alpha+
\frac{1}{2}\int {\tilde\Phi}_\alpha d{\tilde v}\right)^2\\
&&+Q\left[d\theta^2+
c_\theta^2\left(d\psi-\frac{1}{2}(d{\tilde u}-
d{\tilde v})+md{\tilde v}\right)^2+
s_\theta^2\left(d\phi-\frac{1}{2}(d{\tilde u}+
d{\tilde v})-md{\tilde v}\right)^2\right] \,. \nonumber 
\eea
We used the relations
\bea
Q_p=a^2m(m+1),\quad 
\frac{Q_p R_y^2}{Q}=m(m+1),\quad
\eta=\frac{Q}{Q+2Q_p}=\left[1+\frac{2a^2}{Q}m(m+1)\right]^{-1}\rightarrow 1.\nonumber
\eea
Using the time and spatial coordinates, 
${\tilde t}=\frac{1}{2}({\tilde u}+{\tilde v})$, 
${\tilde y}=\frac{1}{2}({\tilde u}-{\tilde v})$,
and the standard formulas for asymptotically AdS spaces, one can extract the charges corresponding to (\ref{AdSDiff}):
\bea\label{DfrPAdS}
P_y&=&
\frac{n_1n_5}{R_y}\left[m(m+1)+\frac{1}{8\pi }
\int_0^{2\pi} d{\tilde y} 
{\tilde f}_\alpha {\tilde f}_\alpha\right]\\
\label{DfrAMAdS}
J_\phi&=&
\frac{n_1n_5}{2}\left[m+1\right],\qquad
J_\psi=
-\frac{n_1n_5}{2}m \,.
\eea
As expected, this agrees with (\ref{sumChrg}).

\end{appendix}

\newpage

\begin{adjustwidth}{-3mm}{-3mm}

%
%
%

\providecommand{\href}[2]{#2}\begingroup\raggedright\endgroup

\end{adjustwidth}

\end{document}